\documentclass[referee]{aa} 
\begin{document}

   \thesaurus{12     
              (12.03.4;  
               12.04.1;  
               12.05.1;  
               12.07.1;  
               12.12.1)} 
   \title{Theory and observations of galactic dark matter}

   \subtitle{}

   \author{Carl H. Gibson
          \inst{1}
          \and
          Rudolph E. Schild\inst{2}
}

   \offprints{C. H. Gibson}

   \institute{Departments of Applied Mechanics and 
Engineering Sciences and\\ Scripps Institution of 
Oceanography, University of California
at San Diego,\\ La Jolla, CA 92093-0411, USA\\
              email: cgibson@ucsd.edu
         \and
             Center for Astrophysics,
    60 Garden Street, Cambridge, MA 02138, USA \\ rschild@cfa.harvard.edu
             }

   \date{Received August 14, 1998; accepted ---}

\authorrunning{Gibson \& Schild}
\titlerunning{Theory and observations of galactic dark matter}

   \maketitle

   \begin{abstract}

Sir James Jeans's (1902 and 1929) linear, acoustic, 
theory of gravitational 
instability gives vast errors for the
structure formation of the early universe.  
Gibson's (1996) nonlinear theory shows that nonacoustic 
density extrema produced by turbulence
are gravitationally unstable at turbulent, viscous, or 
diffusive Schwarz scales 
$L_{ST}, L_{SV}, L_{SD}$, independent of Jeans's acoustic scale $L_J$.
Structure formation began with decelerations of
$10^{46}$ kg protosuperclusters in the hot plasma epoch, 
13\,000 years after the Big Bang, 
when $L_{SV}$ decreased to 
the Hubble (horizon) scale $L_H \equiv ct$, where c is light speed
and t is time, giving
$10^{42}$ kg protogalaxies just before the cooled plasma 
formed neutral H-He gas at 300\,000 years. 
In $10^3$ years this primordial gas condensed to 
$10^{23} - 10^{25}$ kg $L_{SV} - L_{ST}$ scale objects, termed 
``primordial fog particles'' (PFPs), that have 
either aggregated to form stars or 
still persist as micro-brown-dwarf (MBD) galactic ``dark matter.''
Using a precisely measured 1.1-year time delay, Schild (1996) 
suggested from continuous microlensing of quasar Q0957 + 561 A,B
that the mass of the  $10^{42}$ kg lens galaxy 
is dominated by $10^{23} - 10^{25}$ kg
 ``rogue planets ... likely to be the missing mass''. A microlensing
event seen at three observatories confirms
Schild's (1996) claims, and supports
Gibson's (1996) prediction that PFPs comprise most of the dark matter
at galactic scales.  
The Tyson and Fischer (1995) mass profile of a $10^{46}$ kg dense
galaxy cluster suggests a superdiffusive, nonbaryonic, dark matter 
cluster-halo (CH) with $ L_{CH} \approx L_{SD} \gg L_{GH}$. 
Therefore nonbaryonic dark matter should be negligible at galaxy-halo,
$ L_{GH}$, scales.

   \end{abstract}

%

\section{Introduction}
Most cosmologists agree that
99.7 to 99.9 \% of the mass of the universe 
is ``dark matter,'' the smallest part ordinary
``baryonic'' matter (electrons, protons and neutrons)
and the largest part some unidentified, virtually collisionless, 
``nonbaryonic'' material, that behaves like 
(and may just be) neutrinos (\cite{car94}, \cite{mcg98}).  This
view of the world is required for galaxies 
to be bound by gravity with observed rotation rates,
and for the universe to be marginally bound by gravity
(flat) with observed chemical abundances matching the standard hot
Big Bang scenario with inflation (\cite{pbl93,kol94,slk94,gut97}). 

How is it possible for so much matter to be ``missing''
from galaxies and the universe?  We suggest
the ``dark matter'' paradox is mostly the result of 
inappropriate fluid mechanical
theories being applied to  
highly scrambled and turbulent (viscous, diffusive, nonlinear) 
self-gravitational condensation situations
of the actual universe, giving vastly
incorrect predictions for the times, masses, and forms of 
matter structures as they evolved soon after the Big Bang
under the influence of gravity and real fluid effects.  
Unfortunately, the field of
cosmology (\cite{win72,slk89,pbl93,pad93,kol94,slk94} etc.)
has relied exclusively on linear, inviscid, isentropic fluid theory
 (following \cite{jns02}, 1929) by which
self gravitational condensation of a continuous body of
gas with constant density $\rho$ and constant pressure $p$ may occur
only on scales $L_C$ larger than the Jeans acoustic length scale
\begin{equation}
L_C \ge L_J \equiv V_S / (\rho G)^{1/2}
\label{eqa}
\end{equation}
where $V_S$ is the speed of sound in the gas, and $G$ is Newton's
gravitational constant $6.7\,10^{-11}$ $ \rm m^3 \> kg^{-1} \> s^{-2}$.
Jeans's condensation criterion in Eq. (\ref{eqa}) follows
 from a linear perturbation stability
analysis (LPSA) of the hydrodynamic momentum equations, 
with the ``barotropic''
assumption that pressure and density perturbations 
depend only on each other.  Either the LPSA method or the barotropic
assumption will reduce the problem to one of acoustics, without
turbulence, without turbulent mixing, and without
 any validity whatsoever as a description of gravitational
structure formation in cosmology or astrophysics.

Jeans's criterion (\ref{eqa}) forbids condensation
of baryonic matter (by itself) to form such observed structures as galaxies
and stars, but requires gravitational facilitation by 
a previous condensation of nonbaryonic matter 
in the plasma epoch that is impossible according to the
 present fluid mechanical theory.
By hypothesis, cold dark matter (CDM) does not repond to radiation
pressure and consists of nonrelativistically moving massive particles
with small enough $L_J$ to clump and collect galaxy masses of baryonic
gas in gravitational potential wells (\cite{mcg98}).  This idea
fails because nonbaryonic material must be very weakly
collisional or it would have been detected, with a diffusivity
$D_{NB}$ much too large to permit condensation on galaxy scales
because $L_G \ll L_{SD}$, independent of $L_J$ (\S \ref{sect6}).
From the new theory and resulting nonlinear 
cosmology, gravitational decelerations of baryonic matter begin 
in the plasma epoch, first with protosupercluster
and finally galaxy masses, at $L_{SV}$ length scales 
(\cite{gib96}, 1997ab).  The Jeans criterion of equation (\ref{eqa})
is discarded.  Instead, gravitational condensation may
occur only on scales $L_C$ larger than the maximum
Schwarz gravitational condensation scale $L_{SX}$
\begin{equation}
L_C \ge (L_{SX})_{max} \neq L_J ; \> X=V,T,M,D
\label{eqq}
\end{equation}
determined by viscous, turbulent, or magnetic forces,
or by molecular diffusivity 
(\S \ref{sect3}; Eqs. \ref{lsv}, \ref{lst}, \ref{lsm}, \ref{lsd}).  
The idea that radiation
pressure, thermal pressure, or any other kind of
``pressure support'' prevents gravitational condensation
is one further misconception (see \S \ref{press})
associated with the Jeans theory, among many.
   
In the following we review (\S \ref{sect2}) and revise
(\S \ref{sect3}) the linear
theory of self-gravitational condensation.  A very different
scenario for gravitational structure formation emerges at
every stage of cosmological evolution (\S \ref{sect4})
that we term ``nonlinear cosmology''.  
Nonlinear cosmology is as different from linear
cosmology as a turbulent flow is different from a laminar 
flow, and for the same reasons.  Use of
the Jeans theory to describe gravitational condensation
is equivalent to using the linearized Navier Stokes equations to
describe flow of a real fluid at high Reynolds number, where the
actual turbulent flow is vastly different from the
laminar solutions obtained from the linearized equations.   
We focus on the time 300\,000 years after the Big Bang
when the cooling  plasma neutralized to form
 primordial gas.  By the new theory,
the gas rapidly condensed 
at a scale $L_{SV}$ determined by viscous
and gravitational forces to form a ``fog'' of H-He ``particles''
 which still persist
as ``micro-brown-dwarfs'' (MBDs) in galaxy halos 
as the dominant form of 
 dark matter (\cite{gib96}).
The predicted mass of these ``primordial fog particles'' (PFPs) 
 coincides with that of the objects
noted by \cite{sch96}, from  quasar-microlensing observations
in the double image quasar Q0957+561 A,B, as
``rogue planets ... likely to be the missing mass''  
of the quasar lens galaxy.   
New observations for Q0957+561 A,B
 are presented (\S \ref{sect5}) from
three observatories, and all support
 the Schild 1.1-year time delay between images A,B and 
his microlensing interpretion from the image twinkling
frequency.  
In (\S \ref{sect6}) the Tyson and Fischer (1995) 
dense galaxy cluster tomography data 
is interpreted using the new theory to estimate the
diffusivity $D_{NB}$ of the nonbaryonic component of the dark
matter.  We find $D_{NB}$ is trillions of times larger
than $D_{B}$ for baryonic matter, suggesting 
the galactic dark matter for the inner halo 
is mostly baryonic because any nonbaryonic matter 
would diffuse to outer halo or cluster halo ($L_{SD}$)
 scales.  Summary and conclusions are provided in (\S \ref{sect7}).

\section{What is wrong with the Jeans gravitational instability theory?} 
\label{sect2} 

Jeans's theory fails to accurately describe 
gravitational structure formation
because it is linear, whereas gravitational structure formation
is a very nonlinear process.  Linear perturbation
 assumptions (e.g.: $\rho = \rho_0 + \rho '$) 
and the baratropic assumption $p = p(\rho)$
both reduce the problem to one of acoustics, as shown in
detail by Kolb and Turner (1994, p342).  Eulerian equations 
of Newtonian motion describing a perfect fluid 
are assumed, which neglect viscous forces or the gravitational or 
molecular diffusion of density.  By dropping the
convection term $\vec v \cdot \nabla \rho$ in the density
conservation equation, turbulent mixing is precluded.  By dropping
the inertial-vortex force $\vec v \times \vec \omega$
in the momentum equation, turbulence is precluded, where
 $\vec v$ is velocity and $\vec \omega$ is vorticity.  

In the LPSA approach, density $\rho '$, pressure 
$p'$, velocity $\vec v '$, 
and gravitational potential $\phi '$ perturbation
equations, formed by dropping second order
perturbation terms, are cross differentiated with respect to time and space
and combined to give
\begin{equation}
{\partial ^2 \rho ' \over \partial t ^2} - V_S^2 \nabla ^2 \rho '
= 4 \pi G \rho_0 \rho ' ,
\end{equation}
where $\rho '$ is the density perturbation about
the mean density $\rho_0$, 
$t$ is time, and $V_S$ is the
sound speed.  Solutions are of the form
\begin{equation}
\rho ' (\vec r,t) = A \> \rho_0 \> 
exp [-i \vec k \cdot \vec r + i \omega _S t ]  ,
\end{equation}
where $\vec r$ is position, and wavenumber  $\vec k$ and 
frequency $\omega _S$ satisfy the dispersion relation
\begin{equation}
\omega _S ^2 = V_S^2 k^2 - 4 \pi G \rho_0 ,
\end{equation}
with $k \equiv |\vec k |$.  If $\omega _S$ is imaginary, the
mode grows exponentially; if $\omega _S$ is real, the $\rho '$
mode propagates as a sound wave.  
For $k$ less than some critical value, $k_J$,
$\omega_S$ will be imaginary, where
\begin{equation}
k_J = (4 \pi G \rho_0 / V_S^2 )^{1/2},
\end{equation}
giving the Jeans wavenumber criterion
for gravitational instability.  The usual physical 
explanation of this mathematical result is that 
very large wavelength, $\lambda$,
sound waves provide density maxima with
propagation time periods, $\lambda/V_S$, larger than the gravitational
condensation, or ``free-fall'', time, 
$\tau_G \equiv (\rho G)^{-1/2}$, so they
can trigger gravitational condensation, but
shorter waves cannot.  We see below (\S \ref{lic}) that even this
interpretation of the Jeans theory is incorrect.  The
initial condensation scale, $L_{IC}$, is equal to $L_J$,
but its physical basis has nothing to do with acoustics.

Nonacoustic density maxima and minima, particularly the
cold spots and hot spots and helium spots and hydrogen spots 
that must exist in abundance in
the primordial gas due to past or present turbulent
mixing, are gravitationally unstable to condensation
and void formation, but are assumed out of existence 
by the linear Jeans theory.
Reynolds numbers of the primordial gas are well above
critical ($Re_{PG} \approx 10^{9}$) 
so the flow will certainly be turbulent at
the Kolmogorov scale $L_K$ and larger (\cite{gib96}).
Turbulence is defined as an eddy-like state of fluid motion
where the inertial-vortex forces 
$\vec v \times \vec \omega$ of the eddies are larger
than any other forces that tend to damp the eddies (\cite{gib91}).
Turbulence first appears at $L_K \equiv (\nu^3 /\varepsilon)^{1/4}$
and cascades to larger scales by an eddy pairing mechanism, where
$\varepsilon$ is the viscous dissipation rate.
The Reynolds number 
$Re \equiv  ( \vec v \times \vec \omega )/ (\nu \nabla ^2 \vec v)$ 
is the ratio of inertial-vortex to viscous forces,
where $\nu$ is the
kinematic viscosity.  Temperature fluctuations 
$\delta T/T \approx 10^{-5}$ have been detected
in the primordial gas by the COBE 
(COsmic Background Explorer) satellite.  These give 
density gradients that
will be scrambled by the turbulence 
to produce nonacoustic density microstucture which is 
unstable to gravitational condensation and void formation 
at length scales determined by fluid 
mechanical constraints (i.e.: Eq. \ref{eqq}), 
independent of the sound speed $V_S$ or the Jeans scale $L_J$.

Jeans showed that without gravitation, acoustic 
density perturbations propagate
in the continuous gas with sound speed $V_S$, but with gravitation
the speed decreases 
to zero as the wavelength increases to $L_J$ and becomes
imaginary for larger values (\S \ref{jns} Eq. ~\ref{eq1}).  
This result of the Jeans analysis
has been widely interpreted as proof that no condensation 
can occur on scales smaller than $L_J$, as
proposed by Jeans 1902, 1929.  
However, this interpretation is 
incorrect because gravitational condensation on
nonacoustic density maxima to form ``primordial fog
particles'' (PFPs) and gravitational 
expansion of nonacoustic density
minima to form voids was not and could not be 
considered in the linear Jeans analysis,
and cannot be described by LPSA with or without 
the barotropic assumption.  Even when
several terms that Jeans neglected are included and
the barotropic assumption is dropped, the acoustic mode
at the Jeans scale is still the only one that is unstable
(Ned Wright, personal communication 1997). 
By definition, the LPSA method
neglects nonlinear terms of the momentum and density
conservation equations, but these are crucial to
the formation of turbulence and the creation by turbulence
of nonacoustic density extrema, saddle points, and other
zero density gradient configurations that determine the
gravitational creation of structure (\S \ref{nonlinear}). 

Although LPSA has a long history in the fluid dynamics
literature, the technique has generally been supplanted 
for problems involving turbulence and turbulent mixing by 
universal similarity methods, pioneered
by Kolmogorov and Obukhov, involving model-based regimes of 
critical length, time, and scalar field scales and
self-similar, nonlinear, cascades over wide space-time 
ranges (\cite{gib91}).

\subsection{Can anything in the Jeans theory be salvaged?}

According to the new, nonlinear, gravitational
condensation theory, presented here, the Jeans condensation 
criterion is completely irrelevant, and is replaced by new
criteria based on the Schwarz fluid mechanical scales.
The equation for the Jeans scale can be written
\begin{equation}
L_J \equiv V_S / (\rho G)^{1/2} \approx [ (p / \rho ) / \rho G ]^{1/2}
\approx (RT/\rho G)^{1/2}
\label{eq2}
\end{equation}
where $p$ is pressure, $R$ is the gas constant for the particular 
ideal gas or mixture of gases, and $T$ is the temperature.  Let
us examine the two forms on the right.

\subsubsection{Jeans's initial condensation scale $ L_{IC}$} 
\label{lic}

From the
last term of Eq. (\ref{eq2}), the Jeans scale 
$L_J$ can be reinterpreted
as the initial gravitational condensate size $ L_{IC}$ in an 
infinite, motionless gas with uniform $R$, $\rho$, $p$ and  
 $T$ values, so that   
\begin{equation}
L_{IC} \equiv (RT/\rho G)^{1/2} \approx L_J  .
\end{equation}
The fact that the mass of globular clusters
of stars closely matches the Jeans mass at
decoupling is the strongest evidence supporting
the Jeans theory.  However, we see from
the definition of $L_{IC}$ that the physical
basis of the initial condensation mass of the 
primordial gas to form proto-globular-clusters (PGCs)
is an artifact of the ideal gas law $p/\rho = RT$. 
In an isothermal gas this relation permits nonacoustic
density maxima to increase and nonacoustic density minima to decrease 
with precisely compensating pressure changes which 
maintain constant temperature in quasi-equilibrium during
this initial condensation process, so that radiative
heat transfer cannot compensate for the gravitational effects.
This initial condensation instability, and $L_{IC}$, have 
nothing to do with acoustics or Jeans's LPSA theory.

\subsubsection{Jeans's hydrostatic scale $L_{HS}$}

Now consider the penultimate term in Eq. (\ref{eq2}) that includes
the pressure $p$.  
Condensations on the numerous and inevitable 
nonacoustic density maxima within
an initial $L_{IC}$ sized Jeans ``droplet'' are not 
prevented, as often assumed, by 
the constant internal pressure $p$ or constant temperature $T$ 
of the $L_{IC}$ droplet or by its small 
hydrostatic pressure gradients.  
The strongest nonacoustic density maxima are in effect absolutely
unstable with respect to gravitational condensation in a motionless fluid
because their sizes are limited only by molecular 
diffusivity or hydrodynamic forces at the appropriate Schwarz 
scale $(L_{SX})_{max} \ll L_{IC}$ (Eq. \ref{eqq}), as discussed in
the following (\S \ref{sect3}), and immediately begin to collect mass from 
their surroundings by gravitational forces.  As the 
densities of such nonacoustic condensates increase and their sizes
decrease, their individual core internal pressures
respond by increasing to values that decelerate the condensing gas
and eventually provide 
internal hydrostatic equilibrium when their supply of mass ceases
upon the arrival of a void.  Their local Jeans scales 
$L_J$ therefore decrease as an effect of the condensation, 
not as a cause, so $L_J$ in this case should be 
reinterpreted as a hydrostatic
length scale $L_{HS}$; that is,
\begin{equation}
L_{HS} \equiv [ (p / \rho ) / \rho G ]^{1/2} \approx L_{J} 
\end{equation}
where it is assumed that the object is near 
hydrostatic equilibrium surrounded by 
a void with zero pressure.  Again, the hydrostatic
equilibrium process, and $L_{HS}$, have nothing to
do with acoustics or Jeans's LPSA theory.  Neither
have the concepts of ``pressure support'' or ``thermal
support'' commonly used to justify Jeans's theory.

\subsection{Problems with previous attempts to salvage the Jeans theory}

The usual assumptions are incorrect 
(\cite{cha51}, 1961; \cite{bon87}, 1992) that a 
constant internal pressure of a gas
prevents gravitational condensation and that other forces
(turbulence, viscous, magnetic)
simply add additional ``pressures'' which increase the 
minimum condensation scale.  For example, Chandrasekhar (1951) 
proposed a minimum condensation scale
\begin{equation}
L_{JC} \equiv ({V_S}^{2} + {V^2}/2)^{1/2} / (\rho G)^{1/2} \ge L_J
\end{equation}
where $V$ is a characteristic turbulent velocity with a
``turbulence pressure,'' $p_T /\rho ={V^2}/2$.  
One problem with this expression
is that the
``sonic pressure,''  $p_S /\rho = {V_S}^{2}$, 
should be dropped since $p_S$ is constant 
on the scale of the
condensate and larger, 
and a constant pressure
provides no force to resist condensation. 
Other forces prevent 
condensation, but at larger or smaller scales
that are independent of the initial value of 
$L_{IC} \approx L_J$.  Correct
expressions for minimum condensation scales result
when the $p_S /\rho = {V_S}^{2}$ term is not included 
and various expressions for the limiting stresses
divided by density are used.  
The correct ``turbulence pressure'' depends on the
length scale of the condensation, as shown in the
derivation of $L_{ST}$ in Section~\ref{sec1}.  The largest
fluid mechanical stress or the diffusivity of the gas 
determines the initial condensation scale for
a nonacoustic density nucleus, and this scale will be
larger than subsequent local
hydrostatic scales $L_{HS} \approx L_J$.


\subsection{What Jeans thought he was doing}
\label{jns}

Jeans (1929, Chapter XII) was concerned with disputing the ``conjecture'' that
the luminous cores of the ``great nebulae'' might be star-clouds 
rather than gas clouds.  At the time, telescopes could not 
resolve stars in elliptical galaxies or in the cores of spiral galaxies, 
both of which Jeans believed were composed of hot luminous
gas without any stars.  He speculated that the sources of the hot
gas emerging at such spiral galaxy cores might be 
connections to other universes.  
Several arguments were put forth by Jeans 
to prove his erroneous hypotheses, including the 
linear, acoustic, analysis leading to $L_J$.  

Based on his (1902) acoustic gravitational theory,
Jeans (1929, p349) explained that elliptical nebulae
and the cores of spiral nebulae could not condense 
to form stars because $V_S$ would be too large at
the high gas temperatures he assumed must 
exist in these luminous regions; that is,
temperatures giving $L_J$ values larger than the region sizes.  
Only in the spiral arms, which he suggested were
thrown out into the cold by centrifugal forces, could stars
form, since radiative cooling of the spiral arms should reduce 
$V_S$ and $L_J$ to values much smaller
than those in the hot core, and smaller than the sizes of the arms,
so that condensation to form stars could begin.  
Of course we now know that 
elliptical ``nebulae'' and the cores of spiral ``nebulae'' are 
not starless luminous gas but galaxies of $\approx 10^{11}$ 
stars, that spiral galaxy arms are not thrown out by centrifugal
forces, and that the individual
``stars'' Jeans identified in the arms
of spiral nebulae were actually star-burst knots of millions of stars.

  Jeans (1929, p345) assumed
that ``the medium has found a way of acquiring kinetic 
energy in this way indefinitely,'' but since the velocity
of any affected acoustic waves rapidly becomes zero, little or no
increase in kinetic energy could actually result.
Any condensations 
on acoustic scales $L_J$ might generate more 
sound waves on the same scale, 
thus amplifying the Jeans scale condensation effect by resonance.
However, acoustic density maxima become nonacoustic 
density maxima if they accumulate significant quantities of mass by 
gravitational condensation since they will
also accumulate the momentum per unit mass of the medium, which
is zero rather than $V_S$.  
Jeans (1929, p348) himself showed that sound waves do not
propagate with velocity $V_S \approx (dp/d \rho )^{1/2}$ when
the gravitational term is retained in the momentum equation, but
with velocity $V_S ( \lambda )$ given by
\begin{equation}
V_S ( \lambda ) = [ \rho G (\lambda _J ^2 - \lambda ^2)/ \pi]^{1/2},
\label{eq1}
\end{equation}
where $\lambda _J ^2 = \pi V_S ^2 / \rho G = \pi L_J ^2$.  Thus,
self-gravitational acoustic density maxima for
sound waves with wavelengths $\lambda \approx \lambda _J$ are equivalent
to nonacoustic density maxima because 
$V_S ( \lambda _J ) = 0$ from Eq. (\ref{eq1}).

\subsection {What about pressure support and thermal support?}
\label{press}

The Jeans criterion is sometimes derived 
(\cite [p395, Problem 16.14] {shu82}; \cite [p176] {win93}) 
by suggesting that the internal
pressure (or internal energy) of an infinite, homogeneous gas 
can resist gravitational condensation.  
However, a constant internal pressure within a region cannot
resist gravitational condensation in its interior.  
Pressure forces require pressure gradients.
Terms like ``pressure support''
and ``thermal support'' may apply to a star where the external pressure
is negligible, but not to the condensation and void instabilities
existing within a large body
of nearly homogeneous, constant pressure gas 
with embedded nonacoustic density minima and maxima, and
their associated minimum
and maximum saddle points.  At maximum saddle points, large 
gravitational tearing stresses develop to form voids
that enclose the associated density maximum, and at the maximum
point gravitational condensation forces 
monotonically increase with the condensation.    
Such ``pressure support''  and ``thermal support'' 
derivations of $L_J$ are good examples 
of bad dimensional analysis; that is, dimensional
analysis based on an incorrect, misleading, or 
non-existent physical model.  
Setting
the condensation length $L_C$ equal to a function of 
$\rho$, $G$, and $p$ gives
$L_C$ proportional to $(p/\rho)^{1/2} / (\rho G)^{1/2}$, so 
$L_C \approx L_J$ since
$(p/\rho)^{1/2} \approx V_S$.  
What correct physical model is represented?
This hydrostatic analysis only makes sense if the surroundings have 
zero pressure and density and $p$ is the internal hydrostatic pressure.

As discussed previously, $L_{HS} \approx L_J$ 
for any gravitationally bound gaseous object 
can be interpreted as a reflection of its hydrostatic equilibrium, which
ultimately is determined by a complex energy balance.
Hydrostatic equilibrium is established in a spherical gas
blob between pressure gradients and gravitational forces in the radial
direction for this model.  The gravitational force 
\begin{equation}
F_{G} = G \times \rho L^3 \times \rho L^3 / L^2 = \rho ^2 G L^4
\end{equation}
is balanced by the pressure force
\begin{equation}
F_{P} = p \times L^2
\end{equation}
so
\begin{equation}
L_{HydroStatic} \equiv L_{HS} \equiv
[(p/ \rho) / \rho G]^{1/2} \approx V_S / (\rho G)^{1/2} \equiv L_J
\end{equation}
where $F_{G} = F_{P}$ and $V_S$ is the speed of sound.  

Neither internal pressures nor 
hydrostatic pressure gradients
can prevent gravitational condensation on embedded nonacoustic
density nuclei if $L_{ST}$,  $L_{SV}$ and $L_{SD}$ are $\ll L_J$.  On the
other hand, no condensations on scale $L_{J}$ are possible if
$L_{ST}$,  $L_{SV}$ or $L_{SD}$ are $\gg L_J$.
Thus, $L_J$ may overestimate or underestimate
the minimum scale of gravitational
condensation and is therefore quite irrelevant as a gravitational
criterion:  Eq. (\ref{eqa}) is incorrect, obsolete, and misleading,
and is replaced by Eq. (\ref{eqq}).

\section{A new, nonlinear, 
theory of gravitational structure formation}
\label{sect3}

A very different self-gravitational condensation theory and cosmology 
results when condensation is permitted to occur
on nonacoustic density nuclei (\cite{gib88}, 1996, 1997ab); i.e., points 
produced by turbulent mixing of density in the primordial 
gas with maximum and minimum density that are either 
geometrically symmetric and 
therefore nearly stationary with respect to 
the fluid motion, or which move
toward symmetric stationary points 
by molecular diffusion (\cite {gib68}).  According
to the new theory, the criterion for self-gravitational
condensation of density fluctuations on
scale $L_C$ is 
\begin{equation}
L_C \ge (L_{SX})_{max} ; \> X=V,T,M,D
\end{equation}
\label{eqb}
where subscript $X$ of $L_{SX}$
shows whether gravitational condensation is
limited by viscous, turbulent, or magnetic forces,
or by molecular diffusion, respectively (\cite{gib96}).

Density extremum points in great numbers are produced 
by turbulent mixing of temperature, species
concentration and any other fluid property perturbations 
that affect density, and may persist as fossil turbulence long
after the turbulence that produced them 
has disappeared.  The nonlinear advection 
terms $\vec{v}\cdot\vec{\nabla}\rho$ in the density
conservation equation 
\begin{equation}
\frac{\partial \rho}{\partial t}
+\vec{v}\cdot\vec{\nabla}\rho
= [D_{\rho} - L^2 \tau_G] {\nabla}^2\rho
\label{dens}
\end{equation}
and $(\vec{v}\cdot\vec{\nabla})\vec{v} 
= [ \vec \nabla (v^2 /2) - \vec v \times \vec \omega ]$ 
in the momentum conservation equation
\begin{equation}
{\partial \vec v \over \partial t} =
- \nabla (p/\rho + v^2/2) + \vec v \times \vec \omega
+ \nu \nabla ^2 \vec v + \vec F_G + \vec F_M,
\end{equation}
are crucial to the turbulence and turbulent production of 
nonacoustic density extremum
points,
where $D_{\rho}$ is the molecular diffusivity  of density,
$L^2 \tau_G$ is the gravitational diffusivity, $L$ is the
distance to the nucleus, $\tau_{G} \equiv (\rho G)^{-1/2}$
is the gravitational free fall time, 
$\vec F_G$ is the gravitation force, $\vec F_M$
is the magnetic force, and $p$ is pressure.

  Theories like Jeans's that drop the nonlinear terms fail to correctly
describe gravitational condensation in fluids that are 
turbulent or that have been turbulent, with viscous forces and
molecular and gravitational
 diffusivity of density which may affect the smallest
possible gravitational condensation sizes.  Magnetic forces $\vec F_M$
are presumably not important in the primordial gas condensation,
but certainly arise once star formation begins.  

Before nonlinear gravitational condensation 
and void production begins, the density 
diffusivity $D_{\rho}$ simply represents a 
weighted average molecular diffusivity
of temperature and helium concentration, depending on what
linear combination of Fick's law for concentration and Fourier's law 
for temperature dominates the initial small density
fluctuations at scales smaller than $(L_{SX})_{max}$.  
The sources of density for a fluid particle (the right hand side of
Eq. \ref{dens}) are gravity diffusion, thermal diffusion
(cooling), or helium diffusion from the surroundings.  Sources 
and sinks of temperature and helium concentration are neglected, 
so the equation does not apply to the final stages of
condensation (e.g.: in stars) where these may be 
important.  At scales larger
than $(L_{SX})_{max}$, gravitational forces become important
and the effective density diffusivity
$D_{eff} \equiv [D_{\rho} - L^2 \tau_G]$ becomes negative.
Gravity driven ``diffusion'' up density gradients causes density
maxima to grow without limit to form PFPs, and density 
minima to decrease without limit to form voids.  Any CDM clumps
on scales $L \le L_{SD}$ would simply diffuse away.

In natural fluids, density, temperature, and species
concentration extrema are produced by turbulence, and persist
in numerous quantities at Batchelor scales 
$L_B \equiv (D/\gamma)^{1/2}$ of the
turbulence or fossil turbulence (Gibson 1968, 1986, 1988), 
where $D$ is the molecular diffusivity
of such conserved scalar quantities and $\gamma$ is the local rate of strain. 
Fossil turbulence means fluctuations in hydrophysical fields
like density and vorticity produced by turbulence that persist after the
fluid is no longer turbulent at the scale of the fluctuation.  
Most of the density microstructure of the ocean and atmosphere
is fossil density turbulence (Gibson 1986, 1991, 1996).  The Batchelor
scale reflects a local equilibrium between $\gamma$, which produces
more zero gradient points by stretching and compressing 
strong points till they split, 
and $D$ which damps out weak ones and expands the size of strong 
ones to an equilibrium size $L_B$ for all Prandtl number
values $Pr = \nu/D$.  More information about
the kinematics of zero gradient points, lines and minimal gradient
surfaces is given
in Gibson (1968) and in the derivation of the diffusive Schwarz
scale $L_{SD}$ that follows below (Eq. \ref{lsd}).

In the nearly motionless and homogeneous primordial gas, 
the strongest nonacoustic
density minima are absolutely unstable to gravitationally 
driven expansion to form voids, and the strongest nonacoustic maxima
are absolutely unstable to gravitationally driven condensation to form
dense gaseous objects (PFPs), both with hydrodynamically or 
diffusively limited length
scales $(L_{SX})_{max}$ much smaller than $L_J$.
Proto-PFP masses
$(M_{SX})_{min} = [(L_{SX})_{max}]^3 \rho \ll M_J = (L_J)^3 \rho$ are 
$10^{11} \-- 10^{13}$ less
than the Jeans mass $M_J$, contrary to the Jeans theory.
In the chaotic field of numerous nonacoustic density maxima
and minima in competition for mass and space, the strongest density maxima
will dominate and absorb weaker neighbors as they condense
to the maximum density permitted by their mass (contrary
to fragmentation theories of the first 
condensation; e.g., \cite{low76}), just as
the density minima with the smallest density values devour
neighboring weaker minima to form voids with the maximum size
and minimum density permitted by the PFP objects these voids evolve to
surround and contain.

It is crucial to recognize that not all density maxima in fluids
are associated with sound waves.  In natural fluids, 
most $\rho ^{max}$ are not.  Density
depends not only on pressure, but on temperature and species concentration.
As previously discussed, nonacoustic density maxima $\rho ^{max}$ result from 
turbulent scrambling (stirring, mixing, and diffusion) of temperature
and concentration gradients (\cite {gib68,gak88}).
Nonacoustic density maxima move with the fluid velocity if the density
distribution about the maximum is symmetric, or will diffuse with respect
to the fluid toward a position of symmetry and then
move with the fluid velocity.  If the velocity of
the fluid is $\bf \vec v$, the velocity of a density maximum 
${\bf \vec v}_{{\rho}^{max}}$
is given by
\begin{equation}
{\bf \vec v}_{{\rho}^{max}} = {\bf \vec v}  - 
D \left [ 
\left ( \frac {\rho _{,jj1}}{\rho _{,11}} \right) ,
\left ( \frac {\rho _{,jj2}}{\rho _{,22}} \right) ,
\left ( \frac {\rho _{,jj3}}{\rho _{,33}} \right)  
\right ]
= {\bf \vec v} + {\bf \vec v}_{{\rho}^{max}}^{D},
\end{equation}
where the coordinate system is locally aligned with the principal axes of
the tensor $\rho_{,ij}$ with principal values $\rho_{,11}$ etc.,
 $\rho_{,jj}$ represents ${\bf \nabla}^2 \rho$, and $\rho_{,jj1}$ etc.
are the components of the vector ${\bf \nabla} \left ( \rho_{,jj} \right)$.  
One can show from the equation that the direction of 
the diffusive velocity is toward
a position of symmetry where  ${\bf \nabla} \left ( \rho_{,jj} \right) = 0$
and ${\bf \vec v}_{{\rho}^{max}} = {\bf \vec v}$.

The velocity of an isodensity surface ${\bf \vec v}_{\rho}$ with diffusivity D
is given (\cite {gib68}) by
\begin{equation}
{\bf \vec v}_{\rho} = {\bf \vec v} - D 
\left (
\frac {\nabla ^2 \rho}{|\nabla \rho |} 
\right) {\bf \vec s}
= {\bf \vec v} + {\bf \vec v}_{\rho}^{D},
\end{equation}
where ${\bf \vec s}$ is the unit vector of $\nabla \rho$.  Neglecting
convection, a nonacoustic density maximum of scale $L$ will 
grow when the damping diffusive
velocity $|{\bf \vec v}_{{\rho}}^{D}| \approx D/L$ is of the same order
as the velocity of gravitational collapse (on a density
maximum), or expansion (to form a void from a density minimum), 
$L (\rho G)^{1/2}$.  Equality
of the two velocities occurs at a diffusive-gravitational scale termed the 
``diffusive Schwarz radius''
\begin{equation}
L_{SD} \equiv \left ( \frac {D^2}{\rho G} \right) ^ { \frac {1}{4}}
\label{lsd}
\end{equation}
which represents the scale of the smallest nonacoustic density maximum
that can grow by gravitational condensation when the molecular diffusivity
of the density is $D$ and the flow is ``quiescent''.  The dynamical
equivalent of $L_{SD}$ in turbulent mixing theory is the Obukhov-Corrsin
scale $L_{OC} \equiv (D^3/\varepsilon)^{1/4}$, which is obtained (\cite {gib68}) by
setting the diffusive velocity $D/L$ of a scalar field with
diffusivity $D$ equal to the Kolmogorov (turbulence) 
velocity $(\varepsilon L)^{1/3}$ at scale $L$.
For strongly diffusive scalars with 
$Pr \ge 1$, $L_{OC}$ represents the smallest
sized turbulent eddy that can scramble zero 
gradient configurations (ZGCs) from
a region of uniform scalar gradient.

Gravitational condensation of nonbaryonic
fluids, which have enormous diffusivity values $D_{NB}$, 
is prevented by diffusion
for nonacoustic density nuclei, assuming $L_{SD}$ is larger than
Schwarz scales $L_{SV}$ or $L_{ST}$ 
that would otherwise provide the condensation criterion.  This
new length scale is related to the gravitational-inertial-viscous length scale
$L_{GIV} \equiv (\nu ^2 / \rho G)^{1/4} $ (\cite {gib96}) by
\begin{equation}
L_{SD} = L_{GIV} Pr^{-1/2}
\end{equation}
where the generalized Prandtl number $Pr \equiv \nu / D$ (if $D$ is not
the diffusivity of temperature $\alpha$ then the ratio might also be 
called the Schmidt number).  The minimum scale of self-gravitational
condensation is therefore $L_{SD}$ if the flow is 
``quiescent'', where 
\begin{equation}
 Quiescent \> flow \> criteria \> {\rm
[(L_{SX})_{max} = L_{SD}]:} \> \gamma \le (\rho G)^{1/2} 
Pr^{-1} \> {\rm and} \> \varepsilon \le \rho  G D ,
\end{equation}
so that the condensation is limited by diffusive smoothing 
rather than viscous or turbulent forces.  For
gases where diffusion of heat, mass, and momentum occur 
by particle collisions, the diffusivities of these quantities are
all approximately equal, so $Pr \approx 1$ and $L_{SD} \approx L_{GIV}$.

\subsection {Schwarz scale nonacoustic nuclei for non-quiescent flows}
\label {sec1}

Consider an infinite, non-quiescent volume of gas with 
density $\rho$ containing nonacoustic density nuclei; that is,
density maxima $\rho ^{max}$ subject to viscous and turbulence forces 
without compensating pressure fields that cause
them to propagate as sound waves.  
Suppose two density nuclei separated by a distance $L$
have amplitudes $\Delta \rho$ proportional to $\rho$.  The gravitational
force between the nuclei 
\begin{equation}
F_G \approx (G \rho L^3 \rho L^3 )/ L^2 \approx
 G \rho ^2 L^4 
\end{equation}
will be resisted either by viscous forces 
\begin{equation}
F_V \approx
\rho \nu \gamma L^2
\end{equation}
by turbulence inertial-vortex forces 
\begin{equation}
F_T \approx
\rho V(L)^2 L^2 \approx \rho (\varepsilon L)^{2/3} L^2 \approx \rho 
\varepsilon ^{2/3} L^{8/3},
\end{equation}
 where the Kolmogorov 
theory is used to estimate velocity differences,
or by magnetic forces
\begin{equation}
F_M \approx
H^2 L^2
\end{equation}
where $H$ is the magnetic field strength.
According to Kolmogorov's theory the velocity difference between points
separated by a distance $L$ in turbulence with dissipation 
rate $\varepsilon$ is
$V(L) \approx (\varepsilon L)^{1/3}$
if $L$ is larger than the Kolmogorov length scale
$L_K \equiv (\nu ^3 / \varepsilon )^{1/4}$.  According to turbulent
mixing theory (\cite{gib91}), if the magnetic
field $H$ is mixed by turbulence as a passive
scalar, its variance $H^2 \approx \chi_H  \varepsilon ^{-1/3} L^{2/3}$,
where $\chi_H$ is the diffusive dissipation rate of the variance.

Equating the gravitational and viscous forces gives the viscous
Schwarz radius 
\begin{equation}L_{SV} \equiv (\nu \gamma / \rho G)^{1/2},
\label{lsv}
\end{equation} 
equating
the gravitational and turbulence forces gives the turbulent Schwarz
radius 
\begin{equation}
L_{ST} \equiv \varepsilon ^{1/2} / (\rho G)^{3/4},
\label{lst}
\end{equation}
and equating gravitational and magnetic forces gives the magnetic
Schwarz radius
\begin{equation}
L_{SM} \equiv (H^2 / \rho ^2 G)^{1/2} = 
(\chi_H / \varepsilon^{1/3} \rho^2 G)^{3/4}.
\label{lsm}
\end{equation}

A close dynamical similarity exists between $L_{ST}$ and the
Ozmidov length scale $L_R \equiv (\varepsilon/N^3)^{1/2}$ of stratified
turbulence theory (Gibson 1986, 1996).  The Ozmidov scale
of stably stratified turbulence corresponds to the scale
where buoyancy forces $F_B \approx \rho N^2 L^4$ balance inertial-vortex
forces of turbulence $F_I \approx \rho \varepsilon ^{2/3} L^{8/3}$.  In
the self-gravitational case the $ \rm V \ddot{a} is \ddot{a}il \ddot{a}$ frequency
$N = [(g/\rho) \partial \rho /\partial z ]^{1/2}$ 
is replaced by the $N_{\rho} = (\rho G)^{1/2}$, where g is the gravitational
force per unit mass in the $z$ direction (down); so that $L_{ST}$ could
be written $L_{ST} = (\varepsilon/N_{\rho}^3)^{1/2} $.  Just as turbulence
is prevented by buoyancy forces of stable stratification in 
the vertical direction on scales
larger than $L_R$, turbulence is prevented by 
gravitational forces of a density gradient
in space on scales larger than $L_{ST}$ (\cite {gib88}). Condensation
on nonacoustic density nuclei 
takes place at $L_{SV}$ if the Reynolds number is subcritical at the
condensation scale, and at $L_{ST}$ if it is supercritical (\cite {gib96}),
where  $L_{SV} \ge L_{SD} = L_{SV}(\gamma \tau_G Pr)^{-1/2}$ and
$L_{ST} \ge L_{SD} = L_{ST}(\rho G D/ \varepsilon)^{1/2}$ for non-quiescent
flows, independent of $L_J$.

\subsection {How nonlinear gravitational condensation works}
\label{nonlinear}

Figure 1 schematically illustrates the new theory.  The mechanism of
turbulence production of nonacoustic density maxima
and minima is shown on the left.  A uniform
density gradient is distorted by a turbulent convective episode 
(eddy) that ceases (for simplicity) soon after the 
distorted isopycnal surface becomes diffusively unstable and
splits into multiply connected surfaces with maximum points (plus),
minimum points (minus), and associated maximum saddle points and minimum
saddle points (triangles).  Zero gradient density configurations may be characterized
by the signs of the eigenvalues $\vec E$ of the 
density Hessian $\vec{\nabla}\vec{\nabla}\rho$ (\cite {gib68}).  Maximum
points have $sgn\vec E = (---)$, minimum points $(+++)$, maximum
saddle points $(--+)$, minimum saddle points $(++-)$, 
and saddle lines of doublets $(+0-)$. Without gravity, no
other zero gradient configurations are quasi-stable; for example, zero
gradient volumes and minmax saddles with $sgn\vec E$ $(000)$, 
zero gradient surfaces with $sgn\vec E$   $(+00)$ and $(-00)$,
and lines with $sgn\vec E$ $(+0+)$ and $(-0-)$ 
immediately split up into
the quasi-stable configurations.  This might be termed nonlinear
perturbation stability analysis (NLPSA) applied to dynamically passive turbulent 
mixing of scalar fields like temperature (\cite {gib68}), compared
to LPSA, which fails to describe the turbulent mixing
process without gravity just as it fails if gravity is important.

The gravitational force per unit mass $\vec{F_G} = - \vec \nabla \phi $ changes
direction at such zero gradient points, 
where the gravitational potential $\phi$
is determined by density from Poisson's equation
\begin{equation}
\nabla^2\phi = 4\pi G\rho ,
\end{equation}
$\rho$ is density, and $G$ is Newton's gravitational
constant.  Close to density maxima, maximum magnitude 
$\vec{F_G}$ vectors point
toward the maxima, and close to density minima, maximum
$\vec{F_G}$ forces point away.
Along the principle axis of a maximum saddle pointing toward the maximum
the $\vec{F_G}$ forces reverse and act to pull the saddle 
apart to form a void, and succeed at scales larger than $(L_{SX})_{max}$
where other forces and diffusion cannot prevent it 
(Fig. 1, bottom right).  
Gravitational forces per unit area along this axis increase without limit
as the angle between the cones of 
the maximum-saddle point approaches zero, overwhelming pressure
gradients and other forces
that might resist void formation.  

For diffusive decay without gravity or convection, shown at the top right of
Fig. 1, the maximum and minimum saddle points (triangles) merge to form
a minmax saddle  (double triangles) which immediately expands to a saddle line
(filled circles) surrounding the doublet 
where the plane intersects the sphere to form lobes.  Because
$D_{\rho}$ is positive, the decay of the density 
microstructure is monotonically
back toward the original uniform density gradient configuration
without any zero gradient structures.

The evolutions of a density doublet, minimum point-minimum saddle, 
and maximum point-maximum saddle under the influence of gravity and diffusion,
shown at the bottom of Fig. 1, are very different than their decays
with diffusion alone.  Because density
moves up gradient due to gravitational forces, the result is 
similar to what would happen if the sign
of the density diffusivity $D_{\rho}$ (henceforth $D$) were reversed.  
It is assumed that the Jeans scale $L_J$ of the
gas is much larger than the size of the density doublet $L$, which
is larger than the maximum Schwarz scale $(L_{SX})_{max}$ so that
gravitational condensation or void production is not inhibited by
molecular diffusion or hydrodynamic forces.  For the primordial gas
example of interest, the diffusive time scale $\tau_D = L^2/D$ is
 $\approx 30 \, 10^{6}$ years, compared to $\tau_G = (\rho G)^{-1/2}$ of 
$\approx 4 \, 10^{6}$ years or
$\tau_{PFP} = (L_{SX})_{max} / V_S$ about $10^3$ years.  All of the matter
contained in the doublet collapses into the density maximum due to the
imbalance of gravitational forces caused by the doublet, in stages
shown in Fig. 1.  Gravity causes the saddle line to collapse back to
the down-gradient minmax saddle configuration shown,
which is diffusively unstable (\cite {gib68}) but which gravity stabilizes.  
The density maximum monotonically increases
in magnitude and shrinks in size until it is 
completely engulfed by the void, cutting off the mass supply
to the isolated gas particle that continues to collapse to form a PFP.  
The surrounding void propagates
as a rarefaction wave with sonic velocity $V_S$ away from the density minimum
point in all directions driven by radial buoyancy forces, 
forming a nearly spherical cavity that expands until
it reaches the nearest neighboring void.  

Isolated density minima larger than $(L_{SX})_{max}$ form expanding voids
(Fig. 1, bottom right).  Isolated density maxima 
larger than $(L_{SX})_{max}$ are engulfed by expanding voids that
form at their associated maximum saddle points, 
where gravitational forces in opposite directions along the axis toward
the maximum point split the maximum saddle point 
to form the minimum point lobe and saddle line of a doublet.  
As before, the minimum
point lobe of the doublet wraps around the contracting maximum point lobe as the
saddle line moves down gradient, forming
a minmax saddle (double triangles) and an expanding void surrounding the PFP.  
Split maxima larger than $(L_{SX})_{max}$ that
are engulfed by expanding voids form binary PFPs orbiting about
each other in the void.  Numerous other permutations and combinations of these
processes are possible as the continuous primordial gas 
shatters into PFP mass objects in a continuous void due to the
nonlinear gravitational instability of the 
turbulence generated nonacoustic density extrema.

\subsection{Evolution of cold spots, hot spots, and acoustic density extrema}

We can qualitatively demonstrate the absolute gravitational instability of
nonacoustic density maxima and minima to condensation
and void formation, respectively, by considering
the evolution of constant pressure cold spots and hot spots starting
from rest.  
Figure 2 shows schematically the gravitationally driven evolution (left to right)
of a cold spot nonacoustic density maximum in a large body of continuous,
initially motionless gas
with constant initial pressure $p$ (top), 
compared to the evolution of
a hot spot nonacoustic density minimum 
with constant initial pressure $p$ (middle).  
The gravitational instability of both
these nonacoustic extrema is contrasted with the decay of
an acoustic density maximum and its compensating pressure
maximum in the same uniform ambient conditions (bottom).  

The gas is assumed to
be a primordial mixture of hydrogen and helium 
at primordial gas temperatures and pressures so
that the gas is transparent and obeys the perfect gas law.
The initial ``spot'' sizes
$L$ are assumed much smaller than $L_J$
but larger than $(L_{SX})_{max}$.  Initially,
gravitational forces and heat flux vectors are radially away from
the hot spot (left, top) and radially toward the 
cold spot (left, middle), as shown
in Fig. 2, driving monotonically increasing 
velocities $v$ (center to right)
toward the density maximum (top) and 
away from the density minimum (middle).
Because no hydrostatic pressure 
$\delta p$ exists, 
$L_{HS}^{\prime} = 0 \ll L$, as shown (top left) so the
condensation continues.
  
Both the density excess of the cold spot (top) 
and the density deficit of the hot spot (middle) grow until
their differences from ambient temperature $T$ are 
erased (center).  The cold spot is warmed by a combination
of heat transfer and compression, and the hot spot is cooled 
by a combination of heat transfer and expansion.  The pressure
outside the collapsing cold spot core decreases due to the 
increasing inward velocity, from Bernoulli's equation
($p/\rho + v^2/2 \approx constant$). Pressure increases  
near the center of the former cold spot as the velocity goes to
zero and as its density increases from the inward flow of mass, but
$L_{HS}^{\prime} \ll L$ so condensation continues
as shown (top center) in Fig. 2.
Similarly, the pressure decreases in response to the velocity and decreased
density of the former hot spot (middle) as gravity forces from the surrounding
higher density gas pull it to a larger volume.  Heat transfers cease
and reverse as the cold spot and hot spot reach ambient 
temperature, in Fig. 2 (center middle).  Rapid
initial heat transfer does not erase either of the density extrema, it
just spreads them out and may reduce their central density amplitudes
and temperature differences temporarily.  

The constant pressure
initial condition is realistic.  If instead, the density were initially constant 
the pressure would necessarily be below ambient for the cold 
spot and above ambient for the hot spot, and would rapidly approach
the constant pressure initial conditions shown in Fig. 2 (top middle left)
after emission of sound waves, with slight warming of the cold spot and cooling
of the warm spot due to the loss and gain of acoustic energy, respectively.

For the now-isothermal former cold spot (top center Fig. 2), the density
maximum produces a radially inward gravitational 
force $\vec F_G$ and a radially
inward gas velocity $\vec v$ everywhere that 
increase and accelerate indefinitely.  
Compressive work will heat the gravitationally 
condensed gas and hydrostatic
pressures will build up to decelerate the 
condensing gas, but the radius will
monotonically decrease and the mass within any radius, and the 
temperature, will monotonically increase until
either thermonuclear processes begin 
and a star is born, or a nearby growing
void cuts off the supply of mass to form a PFP. The dissipation
rate $\varepsilon = \rho ^{5/6} G^{3/2} M^{2/3}$ corresponding to
star formation with mass 
$M = 10^{29}$ kg is $10^{-11}$ $\rm m^2 \> s^{-3}$,
compared to only 
$\varepsilon \approx 10^{-14}-10^{-15}$ 
$\rm m^2 \> s^{-3}$ in the primordial gas estimated
from COBE measurements (\cite {gib96}).  Therefore, we can
expect PFPs to
form from the primordial gas, rather than small stars.

Radiative heat transfer 
cools the now-hot former-cold spot and increases the rate of 
compaction to form the proto-PFP ( right top), but is not
a necessary condition for the condensation to continue.  
The now-isothermal former hot spot (middle center Fig. 2) is also
absolutely unstable in the homogeneous surrounding
environment shown.  The decreased density produces a 
radially outward gravitational force $F_G$ and a radially
outward gas velocity $v$ everywhere that increase and 
accelerate indefinitely, or until the nearest PFP is encountered.
Radiative heat transfer from the warmer surroundings
heats the now-cold former-hot spot and increases the rate of 
expansion to form the proto-void shown on the 
right.  The expansion is 
accelerated by heat transfer, but the heat transfer is not a necessary
condition for expansion since the void formation accelerates
with or without heat transfer.

The  former cold and hot spots evolve toward the 
``proto-PFP'' and ``proto-void'' configurations, 
as shown on the top and middle
right of Fig. 2.  Both processes are 
monotonic, irreversible, and inexorable.  The collapse
to form a PFP continues until the supply of matter 
vanishes with the arrival at the PFP
of a nearby growing void.  The growth of a void 
ceases when it encounters a PFP.
The PFP is then gravitationally bound with 
length scale $L_{HS}$, and will continue to 
compact as it cools by radiative heat transfer toward ambient temperature.

In contrast, the Jeans acoustic density maximum shown (bottom left) of 
Fig. 2 first collapses
because its compensating pressure maximum expands, and then overshoots to form 
sound waves that propagate radially outward with 
velocity $V_S$.  These move more than a 
wavelength $L$ in a time small compared to $\tau_G$
because $L \ll L_J$ by hypothesis, so their amplitude decreases
as their energy propagates to larger volumes, 
and they eventually dissipate by viscous forces and vanish, as recognized
by Jeans (1902, 1929).  If
any mass is accumulated during the propagation of such acoustic 
density maxima, the
acoustic nuclei will produce nonacoustic density 
maxima and minima because the ambient
momentum of the condensed material is conserved and zero.  
Thus, even if only sonic
perturbations were available in primordial gases, they would trigger
condensation of the gas to PFPs by the same
mechanisms as those illustrated in Fig. 2; that is, self-gravitational 
condensation and void formation triggered by nonacoustic density extrema
on scales $L_J \gg L \approx (L_{SX})_{max}$ 
limited by hydrodynamic forces or diffusion, contrary to the Jeans theory.

\subsection{Evolution of ``cannonball'' and ``vacuum-beachball'' density perturbations}

The isothermal (former) cold and hot spots (center, top and middle) are
functionally equivalent to local ``cannonball'' (mass $M'$) and
``vacuum beachball'' (mass $-M'$) initial density conditions, which are obviously
unstable to gravitational condensation and
void formation in an infinite homogeneous gas at scales independent
of and smaller than $L_J$, contradicting the Jeans criterion
for gravitational instability.  Helium
and hydrogen maximum points have the same effects where they dominate
the density extrema rather than temperature.  Near a helium maximum point
that causes a density maximum larger than $(L_{SX})_{max}$, 
gravitational forces are imbalanced
and cause a monotonic growth in the density to form a proto-PFP
under primordial gas conditions, just as hydrogen concentration
maxima that cause density minima larger than $(L_{SX})_{max}$ will
produce proto-voids.  Again, compressive heating and expansion
cooling cannot prevent the condensation and void formation.  Growths
are accelerated by radiative heat transport but do not require it.  The
momentum equation starting from rest is 
${\partial v_r}/{\partial t} + M' G/r^2 \approx 0$ neglecting magnetic,
viscous, inertial-vortex, and dynamic pressure gradient forces, so
the radial velocity $v_r \approx -M' G t/r^2$. Density changes
occur only near $r=0$, where 
$dM'/dt \approx - 4 \pi r^2 \rho v_r \approx  - 4 \pi \rho M' G t$ 
by the conservation of mass,
so $M'(t) \approx M'(0) exp[2 \pi \rho G t^2]$ increases or decreases
exponentially with $(t/\tau_G)^2$ depending on the sign of the initial density
perturbation $M'(0)$ and the `free-fall' time 
$\tau_G \equiv (\rho G)^{-1/2}$.

\section{Consequences of the new condensation theory: 
nonlinear cosmology}
\label{sect4}

\subsection 
{Plasma epoch deceleration of protosuperclusters to protogalaxies}

During the plasma epoch following the Big Bang,
in the beginning of structure formation,
the radiation dominated speed of sound was
some fraction of the speed of light ($\approx c/3$), 
giving $L_J$ values larger than
the Hubble scale $L_H = ct$.  
Taking $V_S$ = $10^8$ $ \rm m \> s^{-1}$ and $\rho$
$\approx 10^{-18}$ $ \rm kg$ $\rm m^{-3}$ (\cite {win72}) gives 
$L_J \approx 10^8 / (10^{-18} \, 
6.7 \, 10^{-11} )^{1/2} = 1.2 \, 10^{22} $ m, larger than 
$L_H \equiv ct = 3 \, 10^{8} \, 10^{13} = 3 \, 10^{21}$ m, assuming
$t \approx 10^{13}$ s $\approx 300\,000$ y at the plasma to gas transition.  
Therefore, no gravitational deceleration or condensation of baryonic matter
is possible during the plasma epoch by the Jeans criterion
because density information has insufficient time to be transmitted
over scales as large as $L_J$.

After the plasma neutralizes to form gas the Jeans
scale is still very large because the gas is very hot.
Taking $V_S \approx 5.6 \, 10^{3}$ $ \rm m \, s^{-1}$ gives
$L_J \approx 5.6 \, 10^{3} / (10^{-18} \, 
6.7 \, 10^{-11} )^{1/2} = 6.8 \, 10^{17}$ m,
so $M_J  \approx  (6.8 \, 10^{17})^3 \, 10^{-18} =
3.2 \, 10^{35}$ kg, or $1.6 \, 10^{5} M_{\sun}$.  
Thus, structure formation models that rely 
completely on the Jeans criterion and baryonic matter produce
structures much too slowly to match observations.
\cite {pad93} assumes that the hypothetical, 
weakly interacting, nonbaryonic dark matter 
decouples from radiation and begins to condense
during the plasma epoch in order to provide the potential 
wells required to explain the observed baryonic 
matter condensation despite the Jeans criterion. ``Weakly
interacting massive particles'' (WIMPs) provide adjustable
$L_J$ values, by adjusting the speed (hot versus 
cold dark matter) and mass of the ``WIMP''
particles, so that the nonbaryonic dark matter can guide
the baryonic condensation to match observations (\cite {kly97}). Suggested
WIMP particles (\cite {pad93}) have masses in the range
$10^{-38}$ to $10^{-24}$ kg.  ``Axion'' particles
(\cite [p186] {win93}) may be so
numerous that a particle mass as little as $10^{-41}$ kg would supply all the
nonbaryonic dark matter necessary for a flat universe.

However, one problem (among others) with this means of resolving the 
Jeans dilemma is that the nonbaryonic dark matter is 
generically weakly-interacting, similar to neutrinos 
(it may consist entirely of massive neutrinos), 
with detectable collision cross sections 
of order $10^{-40}$ $\rm m^2$ or less (otherwise WIMPs could
be observed).  Such small
collision probabilities imply extremely long mean free
paths $L_{mfp}$ and long times between collisions $\tau_{col}$.
If the $L_{mfp}$ and $\tau_{col}$ values are smaller than the 
space-time scales over which
momentum is to be transferred, then viscous momentum diffusion occurs.
Kinematic viscosities $\nu_{WIMP}$ for ``WIMP'' gases are enormous
compared to $\nu_{baryonic}$ for baryonic gases because
\begin{equation}
\nu \approx L_{mfp} \times V_{particle}
\end{equation}
where momentum is transferred by particle collisions, or
\begin{equation}
\nu \approx \frac {a \times T^4 \times \tau_{col}}{\rho}
\end{equation}
where momentum is transferred by radiation 
(\cite [p57] {win72}). The radiation pressure
is $a \, T^4$, $a$ is the black body constant
$7.56 \, 10^{-16}$ $\rm J \> m^{-3} \> K^{-4}$, and $T$
is the radiation temperature.
Condensation of WIMP dark matter fluids
during the plasma epoch as 
proposed by Padmanabhan (1993) is 
subject to the new viscous-diffusive-gravitational 
condensation criteria of  Eq. (\ref{eqq}), and the
requirement that the condensation scale $L_C$ be less than
the Hubble scale $L_H$
\begin{equation}
 L_H = ct  \geq L_{C} \geq L_{SV} , L_{SD}
\end{equation}
which will not be satisfied early in the plasma epoch
where $t$ is small and $[ \nu \approx D ]_{NB}$ are very large
since $ [L_{SV}, L_{SD}]_{NB}$ for nonbaryonic fluids 
are likely to be greater than $ L_H$ for most of this epoch and 
 much greater than proto-galaxy scales $L_G$ for all of it,
depending on the diffusivity $D_{NB}$ and kinematic
viscosity $\nu_{NB}$ as a function of time.

Little is known of the fluid mechanics of the plasma epoch, except
that temperature, density and viscous diffusivities must have been large
to smooth the temperature and velocities to the small values observed
by the Cosmic Background Explorer (COBE) satellite in the
cosmic microwave background radiation (\cite {slk94}), 
requiring very weak or no
turbulence.  Large viscous dissipation
rates are also suggested by the huge entropy observed, with indicated
initial viscosities $\mu$ of order 
$10^{59}$ $\rm kg \>  m^{-1} \> s^{-1}$ or larger (\cite {bre94})
that produce highly subcritical Reynolds numbers (\cite{gib96}).

Without the Jeans constraint, and without strong turbulence, only large
viscosities or diffusivities can prevent structure 
formation in the plasma epoch (an unfortunate misconception
of linear cosmology is that the expansion of the universe
will damp away all turbulence, whereas expansion actually triggers
and drives turbulence unless constrained by viscosity).  The
magnitude of the viscosity $\nu \approx D$ 
existing then can be inferred from the mass of the largest
structures existing now (superclusters), since these 
presumably were the first objects to form by gravitational deceleration
because they were the first that could possibly form (\cite{gib96}, 1997b).  Supercluster
masses can be estimated because supervoid sizes are known.
The scale of the largest supervoids are observed to be $\approx 10^{24}$ m 
(\cite {kol94}), substantially smaller than the present horizon scale
of  $\approx 10^{26}$ m.  
From the critical density of 
$\rho_{crit} = 10^{-26}$ $\rm kg \> m^{-3}$
and $L_{SV}^3$ we infer a supercluster 
mass $\approx 10^{46}$ kg
that reflects the first deceleration mass when 
$L_{SV}^3 \times \rho$ decreased
to equal the increasing Hubble mass $M_H = (ct)^3 \times \rho$.  
Solutions of Einstein's equations 
(\cite [Table 15.4] {win72}) give the decreasing density $\rho (t)$
of the expanding universe.

Therefore, setting $(ct)^3 \times \rho(t) = 10^{46}$ kg and solving
for t, we find the time when viscous forces first
permitted gravitational structure formation in the universe to be  
$t \approx 4 \, 10^{11}$ s, or 13\,000 y.  
From the density at this time of 
$\rho_{cond}$ $\approx 5 \, 10^{-15}$  $\rm kg \> m^{-3}$
and the rate-of-strain $\gamma \approx t^{-1}$, it follows 
that the kinematic viscosity
$\nu$ was about 1.9 $\, 10^{27}$ 
$\rm m^2 \> s^{-1}$ by setting $L_{SV} \approx
L_H = ct$.  The Reynolds number of the flow $c^2 t / \nu$ is about 20, 
which is below critical, so $L_{SV}$ is the appropriate 
condensation length scale (Eq. \ref{eqq})
assuming it is determined by convection rather than diffusion.  Setting
$\nu \approx D$ gives $L_{SD} \approx 6 \, 10^{19} $  m, 
which is less than
$L_{SV}$, consistent with our assumption that viscous 
forces set the minimum condensation scale.  Such a large
value of $\nu \approx D$ is about $10^2$ larger
than the photon viscosity 
$\nu \approx c/n \sigma_T = 3 \, 10^{8}/ 1.5 \, 10^{11} \, 
0.67 \, 10^{-28} = 3\,10^{25}$ $\rm m^2$ $\rm s^{-1}$ 
expected for Thomson scattering
with the estimated electron density $n$ and Thomson scattering
cross section $\sigma_T$, suggesting the possibility
that the baryonic matter may somehow be coupled to the 
more diffusive nonbaryonic matter.  Since neutrinos appear
to have mass from the Neutrino-98 announcement in Japan, the
Mikheyev-Smirnov-Wolfenstein (MSW) mechanism coupling massive
neutrinos to electrons appears to be a physical possibility worthy
of future study.  The possibility of neutrinos as dark
matter is discussed by \cite{spe97}, p225.

Extrapolation of decreasing $L_{SV}$ with decreasing 
temperature from about $10^5$ to $3\,000$ K gives a decreasing
mass of the decelerations from $10^{46}$ kg
(the mass of a protosupercluster) to about $10^{42}$ kg 
(the mass of a protogalaxy) at the plasma
to gas transition (\cite {gib96}), with a nested foam topology.
The average protosupercluster expansion velocity to date
 is about $10^7$ m $\rm s^{-1}$, compared to 
$10^4$ m $\rm s^{-1}$ for protogalaxies.  The first true
gravitational condensation with increasing density happens
 soon after transition.

\subsection{When the universe turned to fog: the origin of baryonic dark
matter} 

In the present paper, we are primarily interested in a very 
special initial condition 
caused by a very special event, 
where both the condition and event are 
unique in the history of the universe; 
i.e., the vast nearly homogeneous regions of hot (3000 K),
dense ($10^{-18}$ kg $\rm m^{-3}$), 
nearly motionless ($v \le 10^{-5} c$), primordial hydrogen (75 $\%$) and helium
(25 $\%$) gas formed soon after the plasma-gas transition when the photons
of the plasma epoch decoupled from the free electrons of the plasma
as they combined with H and He ions to form neutral gas, and when the 
smallest gravitational
condensation mass permitted by hydrodynamic forces decreased by a factor
of $10^{18}$ from that of a protogalaxy to that of
a PFP.  Soon after that moment, 
nearly all of the baryonic matter 
of the universe turned to primordial fog particles (PFPs).
From our new theory and observations, 
most of these now-frozen, planetary-mass PFPs
still persist invisibly in massive 
galaxy halos as most of the baryonic dark matter.

By our new theory, gravitational condensation 
previous to decoupling was limited
by enormous viscous forces to $L_{SV}$ scales corresponding to
protogalaxies (PGs) which immediately began fragmentation to 
form proto-globular-clusters (PGCs)
at the initial condensation scale 
$L_{IC} \equiv (RT/\rho G)^{1/2} 
\approx L_J$ (\cite {gib96}).  Simultaneous fragmentation
occurred on the nonacoustic density nuclei separated
by Batchelor scales $L_B$, at scales no
smaller than the largest Schwarz scale $(L_{SX})_{max}$. 
The ratio of Jeans to Batchelor scales
\begin{equation}
L_J / L_B = (RT \gamma/ \rho GD)^{1/2}
\end{equation}
is more than $10^4$, substituting known 
primordial $R$, $\rho $, and $D$ values
and assuming a minimum rate-of-strain $\gamma  \approx 1/t$, where 
$t = 10^{13} \> \rm s$ is the age
of the universe at plasma-gas transition (300\,000 years).

Minimum condensation
scales (the Schwarz scales $L_{SX}$) for these nonacoustic nuclei
are determined by fluid mechanical
forces or molecular diffusion, and are also smaller by about $10^4$ than
the Jeans scale $L_J$ for primordial conditions.  Remarkably, this
ancient hydrodynamic state has been measured, and is
known to be extremely quiet, weakly turbulent 
($\delta v /c \ll 0.2$ at horizon scales),
and almost perfectly homogeneous, from 
COBE observations
of the cosmic background radiation at the time of photon decoupling
and plasma-gas transition, so the calculation is possible.
Temperature, and therefore velocity fluctuations, were observed to
be $\delta T/T \approx 10^{-5}  \gg \delta v/c$ on large 
($0.1 \, L_H$) scales (\cite {slk94}).
Because He nucleosynthesis is sensitive to temperature, it seems likely
that fluctuations of He and H concentration may also have
existed at the plasma-gas transition 
as a possible source of nonacoustic density extrema, 
although they have not been detected.   

The initial $M_{IC} = (RT/G)^{3/2} \rho ^{-1/2} \approx 
10^{35}$ kg condensation
to form proto-globular-cluster (PGC) droplets with size 
$L_{IC} \equiv (RT/\rho G)^{1/2} \approx L_J$ 
determined by the
temperature $T$, gas constant $R$ and density $\rho$ of 
this hot primordial gas
is caused by the ideal gas law, not the Jeans scale $L_J$.
According to our present theory, condensations should occur
simultaneously at $(L_{SX})_{max}$ scales within 
PGC droplets on nonacoustic density nuclei with
$(L_{SX})_{max} \approx L \ll L_{IC},$ 
and the condensations are not prevented in any way 
by the approximately constant 
internal pressure and temperature within the PGC droplets
as usually assumed from the Jeans theory.  For the weak
primordial turbulence conditions and the 
homogeneous primordial gas, such
fog-like condensates are estimated to have
moon-planet ($10^{23} -10^{25}$) kg masses (\cite{gib96}).  All of the
baryonic matter of the universe rapidly condensed to form such 
primordial fog particles (PFPs)
by this theory soon after the plasma-gas transition.  Rather than 
requiring the ``free fall'' time 
$\tau_G = (\rho G)^{-1/2} \approx L_J/V_S$ of the primordial
gas of a few million years from the Jeans theory,  
PFPs form in only $L_{PFP}/V_S \approx 10^{3}$ years 
corresponding to the time
for the sonic rarefaction waves of void formation to traverse
average PFP separation distances. 

Therefore, since only a small fraction of the cooling, compacting, and
thus increasingly collisionless 
PFPs in their merging voids are likely to form luminous stars, the baryonic dark
matter of the galaxies consists of any remaining
non-aggregated primordial fog particles. The formation of the first
stars must have taken place soon after the formation of PFPs,
while the PFPs were large and gaseous, 
densely packed, and relatively collisional.  Double stars with
numerous gassy planets should terminate the accretion cascade within PGCs, 
except possibly for a superturbulent gas ball with
growing $L_{ST}$ size at the PG core, rapidly accreting PFPs
but with excellent convective and radiative heat transfer outward.
The small primordial stars of PGCs must 
have formed early if they were to form
at all, perhaps in a few 10 million years, since
the free fall time $\tau_G = (\rho G)^{-1/2}$ increases as the density
of the expanding universe decreases.  By this time
the universe was quite cold, dark, and still quiet, with temperature
$T \approx 30$ K at $t \approx 10^{15}$ s (30 million years) and
with the total mass to luminous mass ratio
$M/L$ nearly infinity.  A few sparks of weakly colliding 
PFPs and PFP agglomerates, all still in  
completely gaseous states well above
the triple point temperature of hydrogen,
might have begun weakly lighting the skies of 
the late dark ages of the universe, along
with a few dimly glowing, gassy brown dwarfs, sparking 
as they collect mass from the $10^5$ PFP 
gas-balls necessary for them to become small stars.
Brilliant gamma ray bursts appear daily as the superturbulence
of distant galaxy cores become relativistic and can no longer
prevent superstar formation, with immediate collapse to
black holes of PGC or multiple PGC mass.  
If no structures had formed in the 
universe by the present time it might be
too late.  The time scale $\tau_G = (\rho G)^{-1/2}$ 
for the present average density of the expanding universe 
($10^8$ less than primordial) exceeds thirty billion years, twice
its estimated age.

Nonacoustic density maxima and minima in the constant pressure, homogeneous
and continuous primordial gas are local regions
that are either colder or hotter than the 
surroundings, and therefore more or less dense.  The strongest 
such hot and cold spot density extrema are 
the most unstable to gravitationally driven 
condensation and expansion and will develop most rapidly.  Radiant and
conductive heat transfer may slow or speed up the condensation and expansion
of such extrema, but not stop them once they start.  

Figure 3 shows the process of primordial-fog-particle (PFP) formation at the
plasma-gas transition based on the present revised theory of self-gravitational
condensation.  On the left is a protogalaxy (PG) droplet of plasma at transition
density $\rho_o \approx 10^{-18}$ to 
$10^{-17}$ $\rm kg \> m^{-3}$ and 
$T_o \approx 3000 \rm K$ at time at $t_o$ (\cite {win72}).
The Jeans scale $L_J$ is larger than the Hubble scale $L_H = ct $, so no condensation
is possible in the plasma epoch by the Jeans theory for baryonic matter, as shown
by the comparison of critical length scales of the process in the upper left hand
corner of Fig. 3.  Cold dark matter theories (\cite {pad93}) 
assume that weakly interacting
massive particle (WIMP) nonbaryonic matter 
decouples from radiation during the plasma epoch, thus
reducing the WIMP sound velocity to values less than $c$, 
where $c$ is the velocity of light, so that
the Jeans scale $L_{J-WIMP} \le L_H$, permitting WIMP
condensations to occur.  These hypothetical 
WIMP condensations then serve to
drive the formation of protogalaxies by providing gravitational
potential wells into which the baryonic matter falls.  However, 
by the present theory and estimates of WIMP diffusivities this
model of galaxy formation is unnecessary and unworkable.
We show evidence in the following that such WIMP 
condensations are prevented by the extremely large generic
diffusivity and long condensation times of the 
WIMP material, so that $L_{SD-WIMP} \gg L_{PG}$, where $L_{SD-WIMP}$
is the Schwarz diffusive scale for WIMP fluid 
and $L_{PG}$ is the size of a protogalaxy
at $t_o$.  Galaxy size ``WIMP-PFPs'' would simply diffuse away. They
are not necessary for protogalaxy formation because the revised
condensation theory predicts 
$L_{PG} \approx L_{HS} \approx L_{SV} \approx L_{ST}$ for plasma at
transition to gas conditions (\cite {gib96}), as shown in Fig. 3. 

The pressure distribution for the gravitationally 
bound protogalaxy (PG) is shown on the left bottom of Fig. 3, 
with pressure  $p_{HS} \approx p_{PG}$ to 
preserve hydrostatic equilibrium of the plasma protogalaxy by radiation
pressure.   
Proto-globular-cluster (PGC) droplets form from the PG droplet at
scale $L_{IC}$, as shown in the center of Fig. 3.   Each of the PGCs then fragment
to form smaller scale PFPs, as shown on the right of Fig. 3, 
according to the condensation and void formation mechanisms described in the last
subsection and shown in Fig. 2.  
The pressures in the proto-PFPs will rapidly become much greater
than the original PGC droplets $p_{PGC}$.  Temperatures of the PFPs are
larger than $T_o$ or $T_{Voids}$, so heat transfer from the PFPs to cooler
regions of space will accelerate their compaction, as shown on the right of Fig. 3.
Because of the quiet initial conditions and the gentle condensation process,
turbulence should continue to be weak, so the hydrostatic Jeans scale
is approximately equal to the Schwarz viscous and turbulent scales
$L_{HS} \approx L_{SV} \approx L_{ST}$ for the proto-PFPs, as shown.  This
means that during the compaction process, viscous, turbulence,
pressure, and gravitational forces are all in approximate balance.

Protosupercluster to protogalaxy mass gravitational
decelerations of the rate of expansion of the plasma universe provide
patterns for subsequent structure formation in the primordial gas.  Within the
protogalaxy $10^{42}$ kg ``droplets,'' further 
decelerations occur on acoustic nuclei with proto-globular-cluster mass
$10^5 M_{\sun}$, reflecting the classical Jeans gravitational instability.  
Only protosupercluster, protocluster, and perhaps protogalaxy 
size structures are within the resolution of the COsmic
Background Explorer satellite (COBE), which shows weak velocity
fluctuations (\cite {bun96}) and therefore weak turbulence 
existed at that time (\cite {gib96}) at scales no 
larger that 1/10 $L_H$.

Within all these structures, condensation on nonacoustic nuclei
should proceed at $L_{SV}$ scales $\ll L_J$, and the entire
baryonic mass of the universe should condense to a ``primordial fog''
of particles of scale $L_{SV}$. 
The mass of such viscous primordial fog particles is given by
\begin{equation}
M_{PFP} = L_{SV}^3 \times \rho = 
\left[ \frac {\mu \gamma} { G}\right]^{\frac {3}{2}} 
\times \frac {1}{\rho ^2} .
\end{equation}
Assuming
T $\approx $ 3000 K, $\gamma = 1/t$, and density
$\rho \approx 10^{-17}$   $\rm kg \> m^{-3}$ (the density of a globular cluster),
gives 
\begin{equation}
M_{PFP} = \left[ \frac {4.4 \, 10^{-5} \, 10^{-13}} 
{ 6.67 \, 10^{-11}} \right] ^{\frac {3} {2}} \,
 \frac {1}  {[10^{-17}]^2}
 = 1.7 \, 10^{23} \> \rm kg,
\end{equation}
where the dynamical viscosity $\mu$ was 
obtained from standard tables.  This moon-mass estimate
of $M_{PFP}$ is a probable lower bound, appropriate
to non-turbulent, $L_{IC}$ scale, 
proto-globular-cluster (PGC) mass ``droplets'' of gas, which is where
the first star formation is most likely to occur.  
Outside the PGC mass ``droplets'' the
density will be less, so that the $M_{PFP}$ masses will be greater.
The average density of the universe at $t=10^{13}$ s was
$\rho \approx 10^{-18}$   $\rm kg \> m^{-3}$ (\cite [Table 15.4]{win72})
giving $M_{PFP} \approx 10^{25}$ kg, or earth mass, $M_{\oplus}$.  

The larger
the mass, the larger the separation distance, and the smaller
the collision probability of PFPs, especially over time as they
become colder and more compact.  protogalaxy droplets with smaller
$\rho$ values, or larger $\gamma$
and $\varepsilon$ values, might form only massive, 
widely separated PFPs that collide too rarely to form stars, leaving
such ``ghost galaxies'' of collisionless ``rogue Jupiters'' forever dark.
A large population of Low Surface Brightness Galaxies 
(LSBGs) have been detected (\cite {bot97}) with properties one might
expect for galaxies with intermediate
``rogue Neptune'' $M_{PFP}$ values barely small enough to aggregate; that is,
LSBGs have small stars, low metal concentrations, low gas concentrations, 
are relatively un-evolved, may be enormous (100 kpc Malin 1 is 30 times the
size of the Milky Way), and 
as a population may have more total
mass than the total mass of conventional galaxies.

Formation of PFPs inhibits formation of turbulence and slows the rate
of structure formation, contrary to standard fragmentation models
of initial star formation (\cite {low76,ree76,slk82}).
PFPs should condense to stars slowly by the weakly-collisional,
clustering cascade of stellar dynamics (\cite {bin87}), since the mass
must increase by a factor of  $10^{6}$ to 
reach $10^{29}$ kg, or $0.1M_{\sun}$,
required for the first
small star to form.  Such clusters are termed ``robust
associations of massive baryonic objects'' (RAMBOs)
by \cite{moo95}, who anticipate severe sampling problems
due to clustering if RAMBOs dominate the halo dark matter
(\cite{gib98}).  

Previous fragmentation models (\cite {slk94})
imply a strongly turbulent
initial star formation process involving all existing gas to produce massive, 
metal-free, stars (termed Population III) and thus a monotonic growth of the
metal/hydrogen ratio since metals form at the expense of hydrogen, but 
neither Population III-stars nor the 
expected metal/hydrogen growth have been observed. 
By our PFP-dark-matter model the metal fraction in stars 
of ancient and recent galaxies should be rather constant 
since star formation, and thus metal production, produces
a proportional amount of primordial material by evaporation of PFPs.
Significant numbers of large, Population III, metal-free 
stars of pure primordial gas never existed by the present 
model.  Formation of ``fog particles'' in the primordial gas
damped out its already weak turbulence.  
Nearly all of the original PFPs should still persist in the 
halos of conventional galaxies as their baryonic dark matter.  
Some galaxies and possibly clusters of galaxies may exist 
with virtually all of their original
PFPs intact, and practically no stars.

\section{A quasar-microlensing event recorded at three observatories}
\label{sect5}

The discussion in the preceding sections leads to the conclusion that
baryonic dark matter dominating galaxy rotation and cluster motions
should consist of a remnant population of ``primordial fog particles'' (PFPs).
These hydrogenous objects should have masses in the 
range $10^{-5} M_{\sun}$ to $10^{-7} M_{\sun}$, and
should be evident in searches for galaxy halo populations by microlensing.
Have the particles been detected?

The search for MAssive Compact Halo Object (MACHO) 
particles in the Halo of our 
Galaxy by microlensing stars of the Large Magelanic Cloud (LMC)
 has resulted in negative but controversial results. 
The classic Alcock et al. (1995abcde) papers of the MACHO 
collaboration usually report lens masses $m \ge 0.1 M_{\sun}$, 
for which event times $t_{sm} \approx 130 \sqrt{m/M_{\sun}}$ (days) are
several months. Star-microlensing 
events by objects in the PFP mass
range last about an hour, and are therefore difficult to detect.
The corresponding TwQSO quasar-microlensing time $t_{qm} \approx 3 \,
10^4 \sqrt{m/M_{\sun}}$ (days) is 9.6 days.

Alcock et al. (1996) report that for a limited subsample of their data,
where several exposures of rapid succession were considered, the low
detection rates indicate non-detection of sufficient mass to make PFP's
the entire mass of a standard spheroidal dark matter Halo. 
Renault et al. (1998) reach the same conclusion from a more intensive
search of a smaller area.  The combined 
MACHO and EROS (Exp\'{e}rience de Recherche d'Objets
Sombres) collaborations (\cite{alc98})
focus on small-planetary-mass objects such as PFPs
in excluding a population with mass 
$M_p = (10^{-7} - 10^{-3}) M_{\sun} $
as more than 25\% of the missing halo mass
within 50 kpc of the Galaxy center (the distance
to the LMC), or
$M_p = (3.7 \, 10^{-7} - 4.5 \,10^{-5}) M_{\sun} $
having more than 10\%.  

However, these exclusions are based on the assumption that
the objects are distributed homogeneously.  This is most
unlikely for such a small-mass population of PFPs, which
are hydrogenous and primordial, and consequently distributed
as a complex array of nested clumps
due to their nonlinear, gravitational-accretion-cascade 
for a wide range of mass to form stars. 
For small $M_p$ values, the number density $n_p$
is likely to become a lognormal random variable with
intermittency factor $I_p \equiv \sigma^2_{ln[n_p]} 
\approx 0.5 ln [M_{\sun}/M_p] = 8.1$
(\cite{gib98}).  For a lognormal random variable, the mean to mode
ratio is $exp[3 I_p / 2] = 1.8 \, 10^{5}$ for $I_p = 8.1$.  
A small number of independent samples of $n_p$
gives an estimate of the mode 
(the most probable value) of a random variable, which is what
is estimated by MACHO/EROS $n_p$ measurements since the LMC occupies
only about 0.1\% of the sky and only a fraction
of this was sampled.  Thus, an exclusion of $10^{-7} M_{\sun}$
objects as $ \le 0.1 M_{halo}$ from an estimate of the mode of $n_p$
is inconclusive, since the mean PFP halo mass 
could be $ \ge 1.8 \, 10^{4} M_{halo}$ from such measurements.
Therefore, we believe that the interpretation of the 
MACHO/EROS statistics is
highly model dependent, and as yet inconclusive.

Results from the quasar-microlensing programs are perhaps more relevant.
Whereas the star-microlensing searches are in a (dimensionless) 
surface mass density realm of 
$10^{-6.5}$, the quasar microlensing optical depth (dimensionless) 
is greater than unity for the TwQSO
system, so the observations are much simpler. A microlensing event or two
should be underway in one of the two observed images at all times.
However this results in some complications of interpretation as well, and
the microlensing program has required the determination of the quasar lens'
time delay. With emerging agreement on a 1.1 year time delay, \cite{sch96}
examined the microlensing presumably originating in the lens galaxy of the
gravitational lens system, and concluded that dominant mass of microlensing
particles had masses around $10^{-5.5} M_{\sun}$. The Schild results suggesting
the detection and identification of the dark matter required confirmation
from independent observations (Sky and Telescope 1996).

A heightened interest in the TwQSO system (Q0957 + 561 A,B) 
resulted from a
prediction by \cite{Kun95} that a rapid decline in the quasar's
brightness should be seen in Feb.---Mar. 1996 in the second arriving B image,
based upon observation of the event in the first arriving A image in
December 1994. Because the time delay was still controversial, with values
of 1.1-years (Schild 1990, Pelt et al. 1995) and 1.4-years (Lehar et al.
1991, Press et al. 1991), it appeared that observations during February and
May 1996 would settle the time delay issue. Thus at least 3 observatories
undertook monitoring programs to observe the predicted event.

At Mount Hopkins, the 15 year monitoring program on the 1.2 m
telescope continued, with observations made by scheduled observers on 109
nights. Four observations were made each night with a Kron-Cousins R
filter, and the observations averaged together for a published nightly mean
brightness value. The quasar brightness was referenced to 5 nearby stars
whose brightnesses were checked relative to each other to ensure stability
of the magnitude zero point. The data are plotted in Figures 4 and 5 of this
report, and data for the first season showing the brightness drop in the
first-arriving A component have been published by \cite{sch97}. 
Data for the second arriving B component in the second year will
be published in our own report about the time delay.

The Princeton data were obtained by \cite{Kun95} using the 3.5m
Apache Point telescope with g and r filters on the Gunn photometric system.
Their data are not published, but data for the first season were posted on
a World-Wide-Web site listed in the Kundic report. We have converted these
data to a standard R filter using the relations given in \cite{ken85}. When
we compare the Princeton results to Mt. Hopkins data for the same dates, we
find an rms residual of 0.029 mag for component A and 0.018 mag for
component B. Unexpectedly, we find the origin of this disagreement not to
be so much in random errors as in a systematic drift in the apparent zero
points in the course of the observing season. Data for the second observing
season have not been presented in tabular form, but a plot of the data
posted at the Princeton WWW site has allowed us to compare the results.
We have taken the Princeton data plot, separated the two colors of data,
and rescaled data for the Mt. Hopkins and Canary Island (Oscoz et al. 1996)
groups to make the comparisons in Figures 4 and 5.

The Canary Island data are posted at the WWW site given in the
report by \cite{osc96}. They were obtained with the 0.8m telescope
using standard R filters and local comparison stars. Because data were
obtained in response to the Princeton challenge of \cite{Kun95},
the Canary Island group reports data for the second season only. We have
compared their data with the Mt. Hopkins data with the assumption that any
Canary Island datum taken within 24 hours 
of a Mt. Hopkins observation had agreeing
dates, and the rms deviations of the two data sets for our 15 agreeing
dates is 0.013 mag for image A and 0.016 mag for B. The error estimates
listed at the WWW web site are considerably larger, averaging 0.022 for A
and 0.020 for B. Because the Canary Island---Mt. Hopkins comparison must
have some error contribution from Mt. Hopkins, it is clear that the posted
Canary Island error estimates are too large by a factor of approximately
2. In our plots of the Canary Island data, Figure 4, we have used the
original posted error estimates.

We show in Figure 4 a comparison of the available photometries for
the first observing season. In the upper plot, the Mt. Hopkins data are
shown with a magnitude scale and zero point for a standard R filter. The
Princeton data have been shown with an arbitrary offset of 0.2 mag. These
magnitudes are determined from a transformation from the Princeton Gunn
g,r photometric system, using the transformation equations determined by
\cite{ken85}. Error bars are shown strictly according to the estimates of
the authors. The data are superimposed in the bottom panel of Figure 4,
and the error bars are suppressed for clarity. It may be seen that the data
agree about as well as predicted from the errors. One artifact that may be
noticed is that there appear to be several points, mostly in the Princeton
data, markedly below the mean trend. It is surprising that these discrepant
points are in the sense of brightness deficiency, because the two principal
error sources, cosmic rays and merging of the two quasar images due to bad
seeing effects, both tend to make the images brighter.
The A component data in the
lower panel will be compared to the observations of image B in Figure 6.

In Figure 5 we show 3 data sets in the upper panel, with their
associated error bars. However as noted previously, we do not actually
have the tabulated data for the Princeton team, and we have scaled the
results of Mt. Hopkins and the Canary Island groups to the Princeton data
as posted in a plot released by the Princeton team.
Thus the plotted magnitudes are on the Gunn photometric system, which
differs from standard R by a zero point offset and a color term of 0.15
mag. In other words, R = r - 0.15(g-r) + Const.  Since image B varied only
from 1.071 to 1.142, a variation of 0.07 mag, we conclude that the scatter
introduced into the comparison of r and R magnitudes has a full amplitude
of 0.01 magnitudes, or a scatter of at most 0.005 magnitudes around a mean
offset. Thus we have simply combined the Princeton r magnitudes with an
arbitrary zero point offset in the comparison with the Mt. Hopkins and
Canary Island R magnitudes in the bottom panel of Figure 5.

We find in Figure 5 (bottom) good evidence that the brightness drop
predicted by \cite{Kun95} did indeed occur at around Julian Date
2\,450\,130. The brightness in the R band did indeed drop almost 0.1
magnitudes, and time delays of 423 days (Oscoz et al. 1997) and 416 days
(Kundic et al. 1997) are determined. However, a remarkable thing happened at
the end of this event, or immediately afterward; a strong microlensing
event was observed, principally in the Mt. Hopkins data. The event may be
seen as a strong downward spike centered on J.D. 2\,450\,151. Although the
event was primarily seen in the Mt. Hopkins data, the brightness did not
recover to the expected level for another 30 days, and for the remainder of
this discussion, we refer to this event as the 3-observatory microlens.

A much better perspective on the 3-observatory microlens comes from
inspection of Figure 6, where we plot data for both observing seasons
combined with the 416-day time delay of Kundic et al. (1997). In this plot
the open and filled symbols refer to the first and second observing
seasons, exactly as in Figures 4 and 5. We consider that from J.D. 2\,450\,151
to 2\,450\,180 the data records are sufficiently discrepant to conclude that
a microlensing event of 30 to 40 days duration and asymmetrical profile
occurred. It is of course possible that more than one event was occurring
at this time. It is likely that another event was seen at 2\,450\,220 $\pm$ 10
days, again seen by 3 observatories. A few other significant discrepancies
may be recognized in this fascinating combined data record, and it is not
surprising that the time delay has been so difficult to determine because
of the influence of this complex pattern of microlensing. On the other
hand, with the time delay now measured, the microlensing provides a
powerful probe of the mass distribution of objects in the lens galaxy, and
perhaps elsewhere (Schild 1996).  The mass of the object causing
the 35 day duration microlensing event at J.D. 2\,450\,151 is $
10^{-6} M_{sun} = 2 \, 10^{24}$ kg.

We now pose the question of the significance level of the detection
of microlensing. We avoid questions 
of \it{a posteriori} \rm statistics by phrasing a
test as follows. A dramatic microlensing event was seen covering dates
J.D. 2\,450\,150-70. During the previous year, a brightness record was
obtained that covered the same time interval. If we average and smooth the
brightness record for the previous year, at what level of statistical
significance can we say each observatory noted a departure in the second
year? Posed this way, we can easily determine that each of the three
observatories observed a departure attributed to microlensing of at least
10 $\sigma$, where the standard deviation
 $\sigma$ has been estimated for the individual data points of
each observatory. Thus we conclude that each of three observatories has
obtained as at least a 10 $\sigma$ result that the second arriving image has
brightness departures attributed to microlensing, because they were not seen
in the first arriving image.

The observation by 3 observatories of the
same microlensing event confirms the conclusion by
\cite{sch96} that low amplitude rapid microlensing
is routinely seen in the Q0957 system.  Interpreted as a
manifestation of microlensing at large optical depth, the
result implies the existence of a large population of masses
in the range $10^{-6} M_{\sun}$, as predicted 
for the initial condensation mass within the primordial gas
(PFPs) in (\S \ref{sect4}).

\section{Nonbaryonic dark matter at galactic scales}
\label{sect6}

What can be said of the nonbaryonic, weakly collisional, matter that
makes up nearly all of the mass of the universe by most cosmologies
(\cite {slk94})?  Tyson and Fischer (1995) have produced the first calibrated
dark matter profile of a dense galaxy cluster, by a
tomographic inversion of 6000 gravitational arcs from 4000 background
galaxies.  The Abell 1689 cluster mass is 
estimated as $10^{45}$ kg, the redshift
$z$ is 0.18, the density is $\rho \approx 5 \, 10^{-21}$ kg/$\rm m^3$,
and the dark/luminous mass ratio is 400 $h^{-1} \approx 800$.
For comparison, average (HST Deep Field) galaxies 
examined by 
the same method  (\cite {del96}) with mass $ \approx 2 \, 10^{42}$ kg
show dark/luminous mass ratios of only $11.4 h \approx 6$
within a radius of $\approx 3 \, 10^{20}$ m (10 kpc).  
The Abell 1689 dark mass distribution 
is smooth and strongly concentrated within a radius of 
$6 \, 10^{21}$ m (200 kpc) from the cluster
center.  Assuming an equilibrium between diffusivity, 
or turbulent or viscous forces,
and gravitational forces gives
a method for estimating the kinematic viscosity of the nonbaryonic
dark matter (WIMP, NB) fluid, by setting the cluster radius 
equal to the maximum of $L_{SD}$, $L_{ST}$, or $L_{SV}$.  For  $L_{SV}$,
\begin{equation}
\nu_{WIMP} \equiv \nu_{NB} \approx L_{SV}^{2} \times \frac {\rho G}{\gamma}
\approx \frac {MG}{L_{SV} \gamma} \approx 
\frac {10^{45} \, 
6.7 \, 10^{-11}}{6 \, 10^{21} \, 3 \, 10^{-18}}
\approx 10^{30} \> \rm m^2 \> s^{-1},
\end{equation}
where $\gamma$ is assumed to be 
that of the universe about 3 By before present.  

Such a large value of $\nu_{NB}$ is consistent with the weakly interacting
nature assumed for the nonbaryonic dark matter fluid.  The Reynolds number
of the flow is about $2 \, 10^{-4}$, so the non-turbulent assumption
would be justified.  
The value $\nu_{NB} \approx 10^{30} \rm m^2 \> s^{-1}$ is 
an upper bound, and much less than 
$c^2 t \approx 10^{34} \> \rm m^2 \> s^{-1}$, which
is the upper bound for physically meaningful 
$\nu$ or $D$ values within the horizon at that time.  
Smaller estimated values
for $\nu_{NB}$ by one or two orders of magnitude 
might be possible based
on the observations, since the effective $\gamma$ for the
embedded galaxies of the cluster could be larger than $1/t$
and $L_{SV}$ indicated by the mass distribution could
also be somewhat larger.
If we assume
$\nu_{NB} \approx D_{NB}$ then we can test 
whether $L_{SD} \le L_{SV}$ as is required.  We find $L_{SD} \approx$
$4 \, 10^{22}$ m (1.3 Mpc) by this method, much larger 
than the size of the core of the cluster 
$6 \, 10^{21}$ m (200 kpc).  Therefore
it appears that diffusion  
rather than viscous forces of 
the expanding universe may
be limiting the size of the cluster to $L_{SD}$.  Substituting
the measured size and density gives 
$D \approx 2 \, 10^{28}  \> \rm m^2 \> s^{-1}$ 
for $D_{NB} \approx \nu_{NB}$.  If turbulence forces, diffusion,
and gravitational forces are all in equilibrium with $L_{SD} \approx L_{ST}$
then $\varepsilon \approx 10^{-3}$ $\rm m^2 \> s^{-3}$.  This value
is close to $\varepsilon \approx V^3 / L$ estimated from dispersion
velocities of $2 \, 10^6$ $\rm m^1 \> s^{-1}$  
(\cite {tyf95}), giving Reynolds numbers near critical 
with $D_{NB} \approx \nu_{NB}$, but it is $10^{11}$ greater than
$\varepsilon \le 10^{-14}$ $\rm m^2 \> s^{-3}$ for
the weak turbulence levels of the universe 
at the plasma-gas transition inferred from the COBE CMR measurements
(\cite {gib96}).

From $D_{NB} \approx m_{NB} / \rho \sigma_{NB}$ it is possible to
calculate an effective collision cross section $\sigma_{NB}$ for the
nonbaryonic fluid if the particle mass $m_{NB}$ can be assumed.  Recent
measurements suggest that neutrinos have mass, which would
be $m_{\nu} \approx 10^{-35}$ kg for neutrinos to produce
 a flat universe.  Substituting this, the 
cluster density $\rho$ and
$D_{NB} \approx 10^{28}  \> \rm m^2 \> s^{-1}$
gives $\sigma_{NB} = m_{\nu}/\rho D_{NB}  
\approx 10^{-43} $ $\rm m^2$ assuming the nonbaryonic
dark matter of the dense cluster consists of such massive neutrinos.
This small cross section supports the assumption that the dense cluster
fluid must be nonbaryonic.

If we accept $D_{NB} \approx \nu_{NB} \approx 10^{28} \> \rm m^2 \> s^{-1}$ 
as representative, it follows that such a large diffusivity
would prevent condensation of most of the nonbaryonic dark matter 
on smaller scale objects like galaxies and could explain why the
observed dark/luminous mass ratio for inner regions of galaxies
(10 kpc) $\approx 6$ is so much smaller
than the ratio for galaxy clusters (200 kpc) $\approx 800$.  
The averaged mass of nonbaryonic dark
matter fluid compared to baryonic mass at galaxy scales 
may be estimated by
\begin{equation}
\frac {M_{NB \> dark \> matter}}{M_{galaxy}} 
\approx \frac {\rho _{universe}}{\rho _{galaxy}}
\approx \frac {10^{-26}}{10^{-20}} = 10^{-6} ,
\end{equation}
which shows nearly all galactic dark matter is probably baryonic.
A contrary view that galaxy halos cannot be baryonic is 
presented by Hegyi and Olive (1986), 
but their arguments and references rely crucially on the 
Jeans gravitational condensation theory and do not consider
the possibility that condensation on nonacoustic nuclei is limited
by $L_{SV}$, $L_{ST}$, $L_{SM}$, or $L_{SD}$ rather than $L_J$.

Figure 7abc summarizes the predictions of the present theory (7a,7c)
of self-gravitational structure formation compared to that
of the Jeans theory (7b).  Viscous forces resist condensation 
in the initial stages of expansion of the hot plasma of the
Big Bang, but some decelerations on protosupercluster to 
protogalaxy scales occur.  A wavy line marks the plasma-gas transition
(photon decoupling) in 7a and 7b.  The Jeans theory in 7b is 
questionable because it relies on nonbaryonic-galaxy-nuclei
condensing in the plasma epoch, which should be prevented by
viscous forces on this super-viscous fluid, 
with $L_{SV-WIMP} > L_{H} \equiv ct$.  Condensation of Jeans mass
proto-globular-cluster droplets within protogalaxy droplets of gas
is shown on the left of 7c, with embedded formations of primordial
fog particles (PFPs).  PFPs become increasingly compact 
and collisionless as they cool, so their aggregation should 
have been slow and gentle to form the
first small stars from growing 
``robust associations of massive baryonic objects'' or 
RAMBOs  rather than rapid and
turbulent to form large, short-lived, stars from turbulent gas 
at scales $L_{ST}$.
Without PFP formation the Jeans mass proto-globular-cluster droplets
of 6c would have been consumed by starbursts and supernova, 
leaving strongly turbulent conditions, with
dense molecular gas clouds and dust clouds typical of spiral galaxy disks,
that would have prevented the formation of the small, ancient,
metal-free, population II stars
observed in dense, spherically-symmetric globular clusters, 
and left large quantities of 
metallic ashes (materials other than H and He) that are not observed.

\section{Summary and conclusions}
\label{sect7}

We conclude that gravitational structure formation 
is generally determined by either diffusion or
viscous, magnetic, or turbulent forces
on nonacoustic density nuclei at diffusive, viscous, magnetic
or turbulent Schwarz scales 
$L_{SD}$, $L_{SV}$, $L_{SM}$ and $L_{ST}$. These scales are
completely independent of the acoustic Jeans scale 
$L_{J}$. An initial condensation scale 
$L_{IC} \equiv (RT/\rho G)^{1/2} \approx L_J$ 
for a much larger body of gas with uniform
pressure and density ratio $p/\rho = RT$.  
This initial condensation process also has nothing to do with the
Jeans gravitational instability theory.  Condensed
bodies of gas in hydrostatic equilibrium have a hydrostatic
scale $L_{HS} \equiv [(p/\rho) /\rho G]^{1/2} \approx  L_{J}$,
but again the relation to the Jeans scale has nothing
to do with acoustics or Jeans's LPSA theory.  $L_{HS}$ is
an effect and not a cause of the condensation.

Jeans's theory of gravitational
condensation fails because it
is linear. Linear perturbation stability analysis
cannot be applied to gravitational condensation 
because it drops the nonlinear inertial vortex
force terms that cause turbulence, and the nonlinear
density convective terms required for turbulent mixing
to form
nonacoustic density nuclei and their associated
zero gradient configurations (ZGCs), such as density
extrema and associated saddle points, which
 actually trigger and guide gravitational
structure formation at gravitational
condensation scales, $L_{SX}$, determined by fluid mechanical
forces or diffusion (\cite{gib96}).  Because
$D_{eff}$ may be negative (\S \ref{sect3}), ZGCs
exhibit complex and remarkable gravitational instabilities
that are quite independent of the Jeans $L_J$ criterion, 
as illustrated in Fig. 1, with rapid structure 
formation at the largest Schwarz scale 
$(L_{SX})_{max}$.      
We show in Fig. 2 that the strongest of
such density extrema in the unique 
continuous gas produced at
the  primordial plasma-gas transition are absolutely
unstable to the formation of proto-PFP condensates and proto-voids on 
length and mass scales determined by the Schwarz condensation
limits $L_{SX}$ and $M_{SX}$, that are smaller than $L_{J}$ and $M_{J}$
by factors of about $10^{4}$ and $10^{12}$, respectively.
Jeans's acoustic criterion (Eq. \ref{eqa})
should be abandoned and replaced by appropriate fluid mechanical
criteria (Eq. \ref{eqq}).

``Cannonball'' and ``vacuum beachball'' nuclei on scales 
$(L_{SX})_{max} \le L \ll L_J$ were introduced
as perhaps the most easily understandable 
counterexamples to Jeans's 1902, 1929
claim that no condensation can take place on scales smaller than
$L_J$ in a homogeneous, motionless, gaseous continuum.  
Using Jeans's initial conditions, these nuclei clearly
trigger condensation and void 
formation by gravity at any scale $(L_{SX})_{max} \le L \ll L_J$,
contrary to the Jeans criterion (Eq. \ref{eqa}) for $L \ll L_J$.
So do any other nonacoustic density nuclei.  
We found in Fig. 2 that cold-maxima
and hot-minima density extrema monotonically condense to form proto-PFPs
and expand to form proto-voids, respectively, with an intermediate state
without heat transfer as they change to hot-maxima and cold-minima that
qualitatively mimic their cannonball-beachball counterparts
as absolutely unstable PFP and void nuclei.

The minimum condensation scale in quiescent fluids,
with $\gamma \le (\rho G)^{1/2} (D /\nu )$ 
and $\varepsilon \le \rho GD$,  is determined by
diffusivity at a new length scale $L_{SD}$ termed the diffusive
Schwarz scale.  Nonbaryonic fluids have enormous
diffusivites $D_{NB}$, and thus enormous $L_{SD}$
values.  Cosmological models based on early
nonbaryonic condensations are strongly affected.  
The new magnetic Schwarz scale, $L_{SM}$, is
probably not important for the electrically neutral condensations of 
greatest interest in the present paper, but 
will be important in stongly ionized supernova 
shocked gas clouds or near galactic nuclei jets, for example.  

By the new self-gravitational condensation
theory, the primordial gas of hydrogen and helium condenses
immediately after the plasma-gas transition to
``primordial fog particles'' (PFPs) within Jeans mass proto-globular
clusters and the protogalaxy
droplets of gas partially formed in the plasma epoch 
along with protoclusters
and protosuperclusters.  The structure formation
sequences for the present theory and the Jeans theory 
are contrasted in Figure 7abc.  Galaxy formation mechanisms
by Jeans's theory fail because diffusivity and
viscous forces prevent condensation of the
nonbaryonic-galaxy-nuclei required to form galaxies 
(\cite {pad93}).  Thus, computer modeling based on cold dark
matter, hot dark matter, and mixed dark matter with n-body interactions
of such nonbaryonic-galaxy-nuclei objects to
match observations of large scale structure are unnecessary
and confuted by the present theory.

According to the $(L_{SX})_{max} \le L \le L_H$ condensation criterion, the
most massive baryonic structures of the universe begin 
forming at a millionth
of the present age rather than a tenth, 
with densities larger by billions and $\tau_G$ values of
millions of years rather than billions.  
If the age of the universe were a year, then
superclusters and supervoids began decelerating at thirty seconds and the
galaxies in an hour as plasma.  At two hours the plasma turned
to gas and the gas turned to fog, but a fog so rarified, and with such
small particles after their freezing at three hours,
that its particles become invisible.  Only with the powerful backlighting 
of a quasar's brightness
and the powerful leverage of a quasar's distance, which together increase
the probability of a PFP microlensing a quasar by a factor of
a million over that for a PFP microlensing a nearby star, has
it been possible to reveal these primordial fog particles, which
have provided the materials for all the stars and which likely comprise
the dominant component of the missing mass of galaxies, at least
within their inner halos (50 kpc).

Just as the twinkling of stars is caused by atmospheric 
density fluctuations, twinkling of quasars is caused by dark
matter objects of galaxy halos.  Any intrinsic variability of a lensed
quasar may be removed by subtraction when 
the time delay between two images is precisely
known, so that the dominant mass of the objects in the halo can be estimated
from the spectrum (\cite {sch96}).
The Schild (1990, see also Schild and Thomson 1997) 
time delay of 1.1-years for the TwQSO quasar is herein reconfirmed by 
data from two more independent observatories.  Two 3-observatory
microlensing episodes were observed, also confirming the
Schild (1996) interpretation that the lens galaxy mass is dominated
by ``rogue planet'' mass objects that we suggest are PFP galactic
dark matter. 

The Tyson and Fischer (1995) calibrated 
measurement of the mass profile of the
dense galaxy cluster Abell 1689 is considered in view of the 
proposed gravitational condensation theory.  By
setting the dark matter condensation scale equal to the diffusive
Schwarz radius $L_{SD}$, the diffusivity 
of the nonbaryonic dark matter fluid $D_{NB}$
is estimated to be $\approx 10^{28}$ $\rm m^2 \> s^{-1}$.  Such a large
value is consistent with the diffusivity and viscosity 
one might expect for weakly-interacting-massive-particle (WIMP) 
nonbaryonic dark matter 
fluids $D_{NB} \approx \nu _{NB}$,
but inconsistent with baryonic $D_{B} \approx  \nu _{B} 
\approx 10^{15}$ $\rm m^2 \> s^{-1}$ values.  
These large
$D_{NB} \approx \nu _{NB}$ values imply 
that only a small fraction of galaxy
halo dark matter is likely to be nonbaryonic for the
inner halo within, say, about 50 kpc of the core, although it may dominate
the outer halo if it extends to $L_{SD}$ scales of $\approx$ 100 kpc.

A long, nonlinear, clustering cascade
is required for PFPs to aggregate by factors of millions in mass
to form stars.  Thus, PFP dark matter will aggregate to
form nested clumps, or RAMBOs (\cite {moo95}), which
become more difficult to detect the longer the cascade.
Probability distributions for random variables X
undergoing self-similar nonlinear cascades over many decades become
highly non-Gaussian, and X becomes
an extremely intermittent lognormal.  This strongly reduces
the probability of microlensing along a single line of sight 
(the optical depth) for PFPs clustering to form stars 
for our random variable X = PFPs per unit volume.
The optical depth for intermittent PFPs (in RAMBOs) 
will be much smaller than for such homogeneously distributed
micro-MACHOs by star-microlensing, so that the average
density of these objects in the halo may be 
underestimated by the star-microlensing method.
Because star-microlensing interpretations 
which exclude small-planetary-mass
objects (PFPs) as the Galaxy halo mass have not accounted
for undersampling errors associated with clumping of such
small objects (\cite{alc98}), we consider such exclusions insecure.

Confirmation by three observatories of the 1.1-year time delay and
PFP-period microlensing with 10 $\sigma$ certainty of TwQSO
quasar-microlensing reported herein gives strong
support to the \cite{sch96} claim that the lens galaxy mass
is ``dominated by rogue planets ... likely to be the missing mass'',
and the \cite{gib96} prediction of the origin of such
objects as ``primordial fog particles''.

\begin{acknowledgements}
We wish to acknowledge advice and discussions with a number of
colleagues, particularly Barnaby Rickett, Paul Libby, Norris Keeler,
Norbert Peters and Peter Bradshaw.  We are
grateful for useful comments of anonymous referees and thank Ned Wright for
providing an extended version of Jeans's linear perturbation analysis.
This work was carried out under the auspices of the Society
for Statistical Geometry.
\end{acknowledgements}

Figure 1.  Schematic illustration of the new turbulent gravitational
structure formation theory.  Nonacoustic density maxima and minima
are produced by a turbulent eddy, shown on the left, which distorts
constant density surfaces of a uniform density gradient until they
become diffusively unstable and break up to form density extrema and associated
saddle points.  Neighboring saddle points (triangles) form a saddle
line (circles) and a doublet.  Without convection or gravitational 
forces the doublet decays
by diffusion and vanishes, as shown top right.  With gravitation
at scales larger than $(L_{SX})_{max}$, the maximum
density lobe contracts to form a PFP and the minimum density lobe 
expands to form a void that surrounds the PFP as the saddle line
contracts to form a minmax 
saddle point (double triangle) down gradient (center).
At the bottom, a minimum point larger than $(L_{SX})_{max}$ forms
a void, and a maximum point larger than $(L_{SX})_{max}$ forms a
doublet, which forms a PFP in a void as before.

Figure 2.  Condensation of a ``cold spot'' nonacoustic density maximum
to form a proto-PFP (top) and expansion
of a ``hot spot'' nonacoustic density minimum to form a proto-void (middle) 
for a continuous, otherwise homogeneous, static, perfect gas representing the
primordial universe after transition from the Big Bang plasma, compared to
the decay of an acoustic density perturbation smaller than the Jeans
scale $L_J$ (bottom) that propagates away and vanishes by viscous dissipation.  
Both these examples of nonacoustic density maxima and minima
initial conditions are absolutely
unstable.  They cause imbalances in the 
initially uniform, static, mass distribution,
producing forces and accelerations toward and away from the nonacoustic density
maximum and minimum (left, top and middle), respectively, with intermediate states 
(center, top and middle) of nearly uniform
temperature where the directions of heat transfer reverse.  After reversal, the
collapse of mass toward the former cold spot accelerates, due to both the increasing
mass within any radius and the increased average density of the proto-PFP
caused by the cooling heat transfer.  The expansion of former hot spot (middle)
is also accelerated for all radii by both the increasing size and the reversed
heat transfer to form the expanding proto-void (right, middle).  
The absolute instability of nonacoustic density maxima and minima
in a homogeneous gas continuum may also be understood by replacing the cold
and hot spot density extrema by a ``cannon ball'' 
and a ``vacuum beach ball'', giving the same 
results as those shown at the right (top, middle), respectively.

Figure 3. Application of the new self-gravitational condensation theory
to the formation of primordial-fog-particles (PFPs) from protogalaxy primordial
gas droplets soon after the plasma-gas transition 300,000 years after
the Big Bang.  

Figure 4. Data for the A (northern) gravitational lens image, recorded in
the October 1994 - June 1995 observing season. In the upper panel,
brightness estimates with error bars are shown as triangles for the
Schild and Thomson (1997) data, and as circles for the Kundic et al.
(1995) data. The R magnitude scale is for the Schild and Thomson data and
the Kundic et al. data is arbitrarily offset 0.2 mag to permit comparison.
In the lower panel, the data are shown superimposed and without error bars,
to show the generally good agreement, especially around the date of the
large quasar brightness drop at 2449715. In Figures 4, 5, and 6 the most
significant 2 digits of the Julian date have been suppressed for clarity.

Figure 5. Data for the B (southern) component recorded in the Nov. 
1995$\--$June 1996 observing season. In the upper panel, filled squares with error
bars are from Oscoz et al. 1996, Triangles are from Schild and Thomson
(1997, in preparation), and circles are from the WWW plot at the site
reported by Kundic et al. (1995).  In the lower panel, data from the three
observatories are plotted without error bars. Generally good agreement is
shown in the comparison, and distinct brightness trends are seen in all
three data sets.

Figure 6. Data from the lower panels of Figures 4 and 5 are shown
superimposed with the same symbol definitions as previously, and for
a 416 day time delay. It may immediately be seen that there is generally
good agreement, and that the second arriving B image (solid symbols)
generally follows the pattern of fluctuation exhibited the year before in
the first-arriving A image (open symbols). However there are important
differences; around Julian Date 2450150-70 a strong brightness drop
occurred that had not been seen in the first arriving A image. Similarly,
around J.D. 2450220 the records differ systematically by several percent.

Figure 7.  Comparison of gravitational structure formation
in the universe for the present theory (7a) 
and the Jeans acoustic theory (7b), and details of galaxy
formation (7c).  The theories diverge in the 
hot plasma epoch (left 7a, 7b) soon after the Big Bang.  By the present 
theory, nonbaryonic (WIMP) matter that dominates the total mass
of the universe is superdiffusive and superviscous 
(see Section 4) with either $L_{SP}$ or $L_{SV}$ values
larger than the Hubble, or causal, scale of the universe $L_H = ct$
during the plasma epoch, preventing condensation to form the
WIMP-galaxy-nuclei assumed and permitted in many cold, hot, and mixed
``dark matter'' theories (\cite {pad93,kly97}) since $L_{J-WIMP} < L_H$.
Such assumptions and theories appear to be questionable, as indicated (7b).
By the present theory (7a), baryonic dark matter is mostly H-He objects 
$\approx 10^{-6} \, M_{\sun}$ that condensed as 
``primordial fog particles'' (PFPs)
soon after the plasma to gas transition (and photon decoupling).  
The Jeans theory permits no baryonic condensation in the plasma epoch,
but condensation of baryonic plasma is permitted 
by the present theory on nonacoustic nuclei when $L_{SV}$ decreases
to $L_H$ or less at $t \approx $ 10,000 years to form a nested-foam
topology of protosupercluster to protogalaxy mass associations
(\cite {gib96}) which turn to PFP fog at decoupling.  The nonbaryonic dark 
matter forms $L_{SD}$ scale superhalos in diffusive-gravitational equilibrium
with the baryonic supercluster and cluster structures as they evolve to form
protogalaxy-droplets of plasma, 
that turn to gas after the wavy line of Fig. 7a.
The detailed evolution of such a protogalaxy-droplet of gas to form 
PGCs, PFPs, RAMBOs, and eventually stars is shown in Fig. 7c.

\end{document}